\documentclass[12pt]{article}
\usepackage{amsmath,amssymb,amsfonts,amscd,amsxtra}
\usepackage{latexsym}

\input{epsf.sty}
\usepackage{epsfig}
\usepackage{graphicx,color}
\usepackage{graphics}
\usepackage{subfigure}
\usepackage{appendix}
\usepackage[top=1in, bottom=1in, left=1in, right=1in]{geometry}

\renewcommand{\theequation}{\thesection.\arabic{equation}}

\renewcommand{\Re}{{\mbox{Re}}}

\newcommand{\norm}[1]{\left\Vert#1\right\Vert}
\newcommand{\abs}[1]{\left\vert#1\right\vert}

\renewcommand\appendix{\par
  \setcounter{section}{0}
  \setcounter{subsection}{0}
  \setcounter{figure}{0}
  \setcounter{table}{0}
  \renewcommand\thesection{Appendix \Alph{section}}
  \renewcommand\theequation{\Alph{section}.\arabic{equation}}
  \renewcommand\thefigure{\Alph{section}.\arabic{figure}}
  \renewcommand\thetable{\Alph{section}.\arabic{table}}
  \renewcommand\thethm{\Alph{section}.\arabic{thm}}
}

\numberwithin{equation}{section}

\date{}

\title{Sensitivity of resonance frequency in the detection of thin layer using nano-slit structures}
\author{
Junshan Lin\thanks{\footnotesize Department of Mathematics and Statistics, Auburn University, Auburn, AL 36849 (jzl0097@
auburn.edu). Junshan Lin was partially supported by the NSF grant DMS-1719851.}
\; Sang-Hyun Oh\thanks{\footnotesize Department of Electrical and Computer Engineering, University of Minnesota, Minneapolis, Minnesota 55455, USA.}
 \; and Hai Zhang\thanks{\footnotesize 
 Department of Mathematics, 
  HKUST,  Clear Water Bay, Kowloon, Hong Kong (haizhang@ust.hk). Hai Zhang was supported by HK RGC grant GRF 16304517 and GRF 16306318.}
  }

\begin{document}
\maketitle
\textbf{Abstract} We derive the formulas for the resonance frequencies and their sensitivity when the nano-slit structures are used in the detection of thin layers.  
For a thin layer with a thickness of $H$ deposited over the nanostructure,
we show quantitatively that for both single and periodic slit structures with slit aperture size $\delta$, 
the sensitivity of resonance frequency reduces as $H$ increases. Specifically, the sensitivity is of order $O(\delta/H)$ if $H >\delta$ and of order $O(1+\ln H/\delta)$ otherwise.
The evanescent wave modes are present along the interface between the thin dielectric film and ambient medium above.
From the mathematical derivations, it is observed that  the sensitivity of the resonance frequency 
highly depends on the effect of evanescent wave modes on the tiny slit apertures.

\setcounter{equation}{0}
\setlength{\arraycolsep}{0.25em}

\section{Introduction}
The resonances induced by subwavelength holes etched in a slab of noble metal
can trigger the so-called  extraordinary optical transmission (EOT) \cite{Ebbesen-1998, Vidal-2010}.
This allows for the detection of a variety of biomolecular events in a label-free and highly sensitive manner from the shifts of  
sharp transmission peaks. We refer the reader to \cite{Brolo, Blanchard-Meunier, Cetin, Gome-Crutz, Li, Oh_Altug, Rodrigo, Soler} and references therein
for detailed discussions.
One fundamental question in such applications is the sensitivity of resonance frequencies, or equivalently, how the transmission peaks shift 
with respect to the refractive index change or the profile change of the biochemical samples.
There have been both experimental and numerical studies concerning the resonance spectral sensitivity \cite{Brolo, Dahlin, Lee}.
However, so far rigorous analytical formulas have been limited to the bulk sensitivity only when
the refractive index of the entire surrounding medium changes; see, for instance, \cite{PHSF}. In this paper, we aim to derive analytically the resonances frequencies
and their sensitivity for the thin film detection, where refractive index only changes locally over the surface of nanostructures.
This occurs in many realistic biosensing applications such as detection of molecular binding and bacterial infections, analysis of live cell secretion, etc.
The mathematical derivations are built upon our recent work on the quantitative analysis of the EOT phenomenon for various nanostructures in \cite{LONR} - \cite{lin_zhang19_3}. 

We point out that other than subwavelength hole structures, plasmonic nanoparticles have also been widely used for bio-chemical sensing,
due to their ability to trigger localized surface plasmon resonance (LSPR) as well as strong electromagnetic field enhancement.
In the presence of local changes in the environment induced by surface-bound molecules, the corresponding resonant frequencies will also be shifted, which encodes the information about the local refractive index changes. We refer the readers to the review papers \cite{Anker2008, Willets2007} and the references therein for nanoparticle-based LSPR biosensors, and \cite{Ammari-Yu-2018, Ammari-zhang18-1, Ammari-zhang18-2} for the mathematical treatment of sensing by nanoparticles.

\begin{figure}[!htbp]
\begin{center}
\includegraphics[height=4cm]{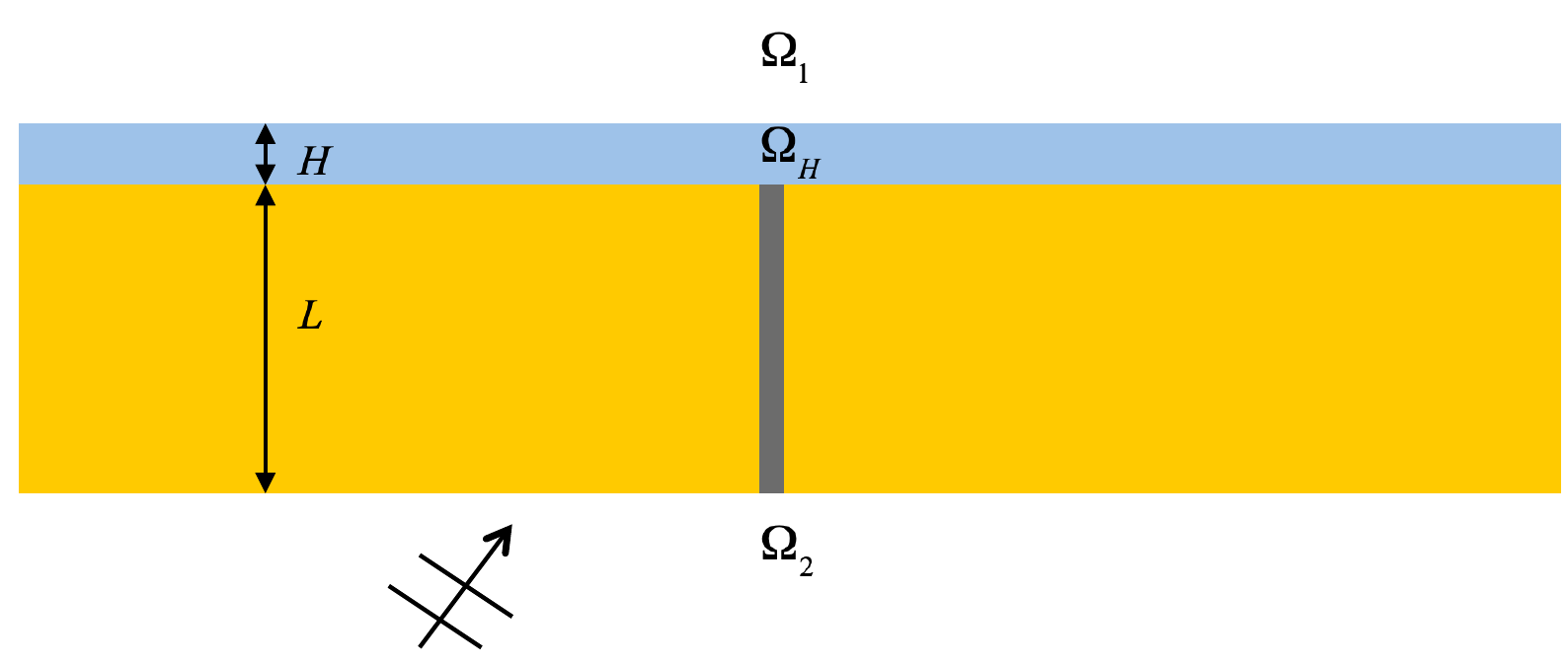}  
\includegraphics[height=4cm]{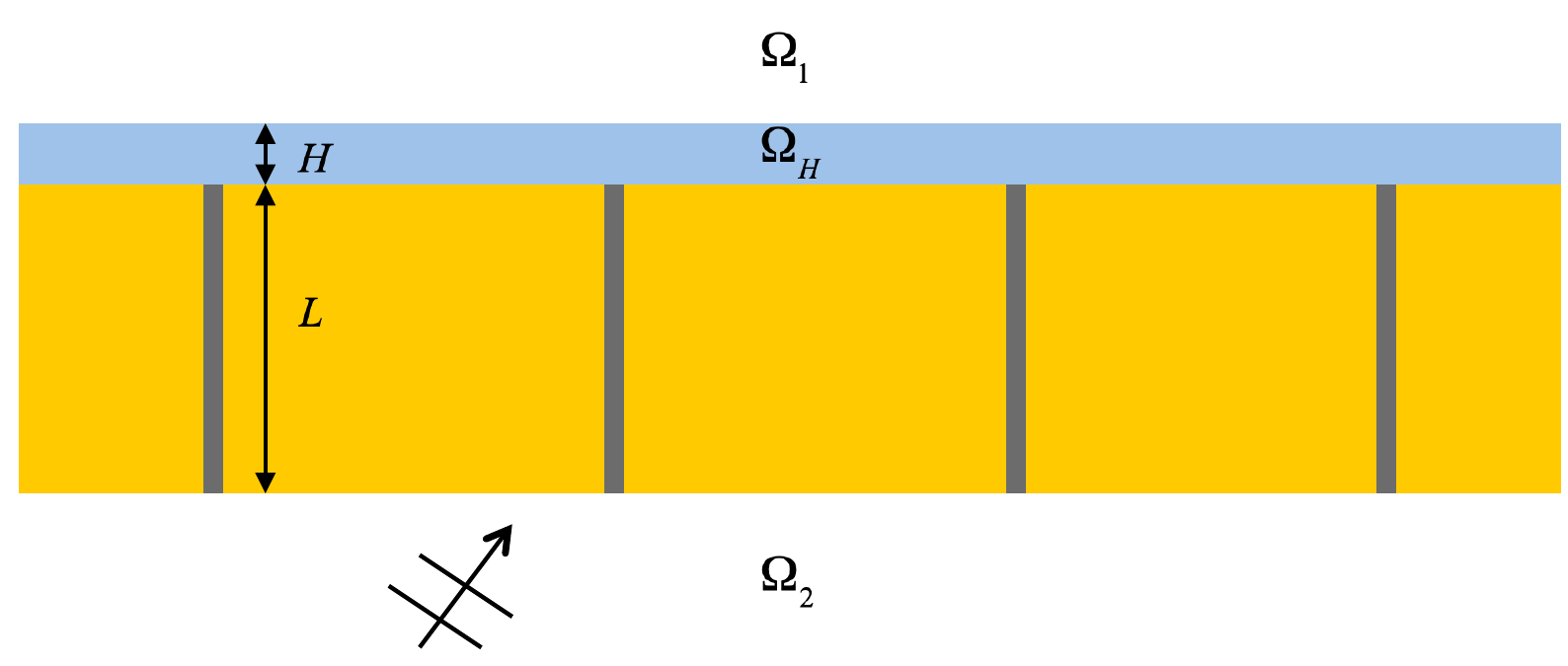}
\caption{Setup for the sensing of a thin layer using a single nano-slit or a peirodic array of nano-slits. }\label{fig:prob_geo}
\end{center}
\end{figure}

In this paper, we consider the configuration where a thin dielectric film (as a simplified model for surface-adsorbed molecules) is deposited on top of the single nano-slit structure or periodic structure as shown in Figure \ref{fig:prob_geo}.
The thin layer attains a thickness of $H$  and occupies the domain $\Omega_H$.
With a suitable scaling of the mathematical model, we may assume that 
each of the rectangular nano slits arranged in the metallic slab attains the height $L=1$ and the width $\delta \ll 1$.
In the idealized case where the metal is a perfect conductor, and without the deposited thin layer, it was rigorous established in \cite{lin_zhang17, lin_zhang18_1, lin_zhang18_2} that the slit structure in both settings (single slit and periodic slits), can induce Fabry-Perot type resonances. The resonant frequencies correspond to the peak values of the spectral transmission line, which is measured by spectrometers. 
In the more realistic configuration when the 
metal is not perfect conducting, the electromagnetic fields can penetrate through the metals and this may induce shift or widening  
to the Fabry-Perot resonances depending on the skin depth. The associated analysis is much more complicated. We refer \cite{lin_zhang19_1, lin_zhang19_2, lin_zhang19_3} for the mathematical and numerical treatment in this direction. 
In this paper, we will focus on the case when the metal is perfect conducting. This is a good starting point to allow us to focus on the main ideas of sensing using the resonant structures. With the presence of the thin layer, the Fabry-Perot resonance frequencies obtained from the transmission data will be shifted due to the near field interactions of the slit structure and the dielectric material in the thin layer. The shift depends on the permittivity and the thickness of the thin layer. In typical sensing applications, one is interested in monitoring the change of thickness from the measured spectral shift, which corresponds to finding the explicit dependence of the shift on the thickness and its sensitivity, as is investigated here.

To be more specific, we assume that the relative permittivity $\varepsilon(x)$  takes the value of $\varepsilon_s$ and $\varepsilon_\ell$ in the slits and the thin layer respectively.
The corresponding refractive indices are given by $n_s = \sqrt{\varepsilon_s}$ and $n_\ell = \sqrt{\varepsilon_\ell}$.
The ambient medium is assumed to be vacuum with the relative permittivity value $\varepsilon(x)=\varepsilon_0=1$. 
We also set the ratios
\begin{equation}\label{eta}
\eta_1:=\varepsilon_\ell/\varepsilon_s \quad \mbox{and}  \quad \eta_2:=\varepsilon_0/\varepsilon_s
\end{equation}
throughout the paper. Let us consider the transverse magnetic (TM) polarized scattering when a time-harmonic electromagnetic wave impinges from below the slab
($\Omega_{2}$ in Figure \ref{fig:prob_geo}) .
The third compoponent of the incident magnetic field is given by
$u^\mathrm{inc} = e^{i k( x_1\sin\theta \,+\, x_2\cos \theta )}$ 
in which $k$ is free-space wavenumber, and $\theta$ is the incident angle. 
The total field $u$ consists of the incident field $u^\mathrm{inc}$ and the diffracted field $u^\mathrm{diff}$ in the lower domain $\Omega_2$,
and only the diffracted field $u^\mathrm{diff}$ in the upper domain $\Omega_1$ above the thin layer.
Let $\Omega_\delta$ be the slit region and $\Gamma_\delta$ be the union of the slit apertures. 
We denote by $\Omega$ be the region exteior to the metal, which consisits of the slit domain $\Omega_\delta$,
the top and bottom domains $\Omega_1$,
$\Omega_2$, and $\Omega_H$.
Then for both configurations, the total field satisfies
\begin{equation} \label{eq-scattering1}
\left\{
\begin{array}{llll}
\vspace*{0.1cm}
\nabla \cdot \left(\dfrac{1}{\varepsilon(x)} u \right) + k^2 u = 0   \quad\quad  \mbox{in} \; \Omega, \\
\vspace*{0.1cm}
\dfrac{\partial u}{\partial\nu} = 0  \quad \mbox{on} \; \partial  \Omega, \\
\vspace*{0.1cm}
[u] = 0,  \left[\dfrac{1}{\varepsilon} \dfrac{\partial u}{\partial \nu}\right] = 0    \quad\quad  \mbox{on} \; \partial\Omega_1 \cup \Gamma_\delta.
\end{array}
\right.
\end{equation}
In the above, $[\cdot]$ denotes the jump of the quantity when the limit is taken along the positive and negative unit normal direction $\nu$.
In addition, the diffracted field $u^\mathrm{diff}$ satisfies outgoing radiation conditions at infinity,
which will be enforced naturally by the Green's functions in the integral formulations in this paper.

Our goal is to provide analytical formulas for the resonance frequencies in both configurations and investigate their
 sensitivity with respect to the change of the layer thickness $H$.
Asymptotic expansions of resonances are obtained via an equivalent boundary integral equation formulation for the scattering problem \eqref{eq-scattering1}
and its asymptotic analysis. The sensitivity analysis then boils down to the perturbation analysis of the underlying layered Green's function.
The studies for the single nano-slit and the periodic nano-slits case will be carried out in Section 2 and 3 respectively.

\setcounter{equation}{0}

\section{Sensitivity of resonance frequency for the single nano-slit configuration}
\subsection{Boundary integral formulation and asymptotic analysis} \label{sec-formulation}
We choose the origin of the coordinate to be the lower left corner of the slit so that
the upper and lower slit aperture can be expressed as
$\Gamma_{1,\delta} :=\{ (x_1,1) \;; 0<x_1<\delta \}$ and $\Gamma_{2,\delta} :=\{ (x_1,0) \;; 0<x_1<\delta \}$
respectively.
Let $ g_1(x,y)$ be the layered Green's function in $\Omega_1\cup\Omega_H$ with the Neumann boundary condition along metallic slab boundary.
Applying the Green's formula in $\Omega_H$ gives 
\footnote{For a function $\varphi(x)$ defined in $\Omega$, $\varphi(x_+)$ and $\varphi(x_-)$ denotes the limit of the function when $x$ approaches the 
aperture from the above and below respectively.}
$$u(x) = \int_{\Gamma_{1,\delta}} g_1(x,y) \dfrac{\partial u(y_+)}{\partial \nu} ds_y, \quad x\in\Omega_H.  $$
Similarly, using the Green's function $g_2(x,y)$ in $\Omega_2$ with the Neumann boundary condition, one obtains
$$u(x) = \int_{\Gamma_{2,\delta}} g_2(x,y) \dfrac{\partial u(y_-)}{\partial \nu} ds_y + u^\mathrm{inc}(x)+u^\mathrm{ref}(x), \quad x\in\Omega_2,  $$
where $u^\mathrm{refl}(x)= e^{i k( x_1\sin\theta \,-\, x_2\cos \theta )}$ is the reflected field of the lower slab boundary at the absence of the slit.
The solution inside the slit can be expressed as
$$ u(x) = -\int_{\Gamma_{1,\delta}} g_i(x,y) \dfrac{\partial u (y_-)}{\partial \nu} ds_y - 
\int_{\Gamma_{2,\delta}} g_i(x,y) \dfrac{\partial u (y_+)}{\partial \nu} ds_y\quad \mbox{for} \; x\in \Omega_\delta, $$ 
in which $g_i(x,y)$ is the Green's function inside the slit $\Omega_\delta$ with Neumann boundary condition.
Therefore, by taking the limit of the above integral to the slit apertures and 
imposing the continuity condition of the electromagnetic field,
we obtain the following system of boundary integral equations for 
$\tilde\varphi_1 := \left. \frac{\partial u (y_-)}{\partial \nu} \right|_{\Gamma_{1,\delta}} $ and 
$\tilde\varphi_2 := \left. \frac{\partial u (y_+)}{\partial \nu} \right|_{\Gamma_{2,\delta}} $:
\begin{equation} \label{eq-scattering2} \nonumber
(2.4) \left\{
\begin{array}{ll}
& \hspace*{-10pt} \eta_1 \displaystyle{\int_{\Gamma_{1,\delta}}}  g_1(x,y) \tilde\varphi_1(y) ds_y  + \sum_{j=1}^2\int_{\Gamma_{j,\delta}} g_i(x,y) \tilde\varphi_j(y)  ds_y
=0\quad \mbox{on} \,\, \Gamma_{1,\delta}, \\
\\
& \hspace*{-10pt} \eta_2  \displaystyle{\int_{\Gamma_{2,\delta}}}  g_2(x,y) \tilde\varphi_2(y) ds_y  + \sum_{j=1}^2\int_{\Gamma_{j,\delta}} g_i(x,y) \tilde\varphi_j(y)  ds_y
+ u^\mathrm{inc} + u^\mathrm{ref}
=0 \quad \mbox{on} \,\, \Gamma_{2,\delta}. 
\end{array}
\right.
\end{equation}
Recall that the parameters  $\eta_1$ and $\eta_2$ are defined by $\eta_1=\varepsilon_\ell/\varepsilon_s$ and $\eta_2=\varepsilon_0/\varepsilon_s$ in \eqref{eta}.
Moreover, due to the symmetry of the structure, the slit Green's function satisfies the following over the slit apertures:
$$ g_i(x_1, 1; y_1, 1 ) = g_i(x_1, 0; y_1, 0)  \quad\mbox{and} \quad g_i(x_1, 1; y_1, 0 ) = g_i(x_1, 0; y_1, 1). $$

If one rescales the slit aperture to an interval of size $1$ by a change of the variable
$x=\delta X$ and $y=\delta Y$, and introduce the following quantities for $X \in (0,1)$:
\begin{eqnarray*}
&& \varphi_1(X) := \tilde\varphi_1(\delta X), \varphi_2(X) := \tilde\varphi_2(\delta X),  f(X):= u^\mathrm{inc}(\delta X, 0)+u^\mathrm{ref}(\delta X, 0), \\
&& G_1(X, Y) :=  g_1(\delta X, 1; \delta Y, 1), \quad G_2(X, Y) :=  g_2(\delta X, 0; \delta Y, 0);   \\
&& G_i(X, Y) := g_i(\delta X, 0; \delta Y, 0 ), \quad \tilde G_i(X, Y) := g_i(\delta X, 1; \delta  Y, 0 ),
\end{eqnarray*}
the above system of integral equation reads
\begin{equation}\label{eq-scattering3}
\left[
\begin{array}{cc}
\eta_1 T_1+T^i    & \tilde{T}^i \\
\tilde{T}^i  & \eta_2 T_2+T^i 
\end{array}
\right] \left[
\begin{array}{llll}
\varphi_1    \\
\varphi_2 
\end{array}
\right]=\left[
\begin{array}{llll}
0  \\
f/\delta  
\end{array}
\right],
\end{equation}
where $T_1$, $T_2$, $T^i$ and $\tilde T^i$ are integral operators defined for $X\in (0,1)$ with the kernels $G_1(X, Y)$, $G_2(X, Y)$,
$G_i(X, Y)$ and $\tilde G_i(X, Y)$ respectively.

A scattering resonance of \eqref{eq-scattering1} refers to a complex-valued $k$
such that \eqref{eq-scattering1} attains a nontrivial solution when the incident field is zero. 
The real part of $k$ is called resonance frequency, which is shifted when the thickness of thin film $H$ changes.
From the above discussions, this boils down to solving for
the characteristic values of the operator integral operators in \eqref{eq-scattering3} when $f=0$.
To this end, we perform the asymptotic analysis of the integral operators for $\delta \ll 1$.

The explicit expressions of the Green's functions $G_1$ and $G_2$ are given by
\begin{equation}\label{eq:G1G2}
G_1(X, Y) = -\frac{i}{2} H_0^{(1)}(kn_\ell|X-Y|) + G_H(k;X,Y), \quad G_2(X, Y) = -\frac{i}{2} H_0^{(1)}(k|X-Y|),
\end{equation}
where $H_0^{(1)}$ is the zero order Hankel's function of the first type, and
\begin{eqnarray}\label{eq:GH}
G_H(k;X,Y) &=& -\frac{2i}{\pi}\int_0^\infty \dfrac{ A_- (k,\xi)  e^{i2\rho(kn_\ell,\xi) H }  \cos(\delta\xi(X-Y)) }
{\rho(kn_\ell,\xi)  \big( A_+ (k,\xi) - A_- (k,\xi) e^{i2\rho(kn_\ell,\xi) H } \big )    }\, d\xi,  \label{eq:GH}  \\
A_\pm (k,\xi) &=& \rho(kn_\ell,\xi) \pm \rho(k, \xi) \varepsilon_\ell, \quad \rho(k,\xi) = \sqrt{k^2-\xi^2}.   \label{Apm}
\end{eqnarray}
The slit Green's function are given by
\begin{eqnarray}\label{eq:green}
&& G^i(X, Y) = \sum_{m,n=0}^\infty\dfrac{c_{mn}\alpha_{mn}}{\delta}\cos(m\pi X) \cos(m\pi Y), \\
&& \tilde G^i(X, Y) =\sum_{m,n=0}^\infty\dfrac{ (-1)^n c_{mn} \alpha_{mn}}{\delta}\cos(m\pi X) \cos(m\pi Y),
\end{eqnarray}
where
\begin{equation*}
c_{mn}=\dfrac{1}{k^2\varepsilon_s-(m\pi/\delta)^2 - (n\pi)^2}, \quad
\alpha_{mn} = \left\{
\begin{array}{llll}
1  & m=n=0, \\
2  & m=0, n\ge 1 \quad \mbox{or} \quad n=0, m\ge 1, \\
4  & m\ge 1, n \ge 1.
\end{array}
\right.
\end{equation*}
Using the expansion of the Hankel function, it follows that (cf. Lemma 3.1 in \cite{lin_zhang17})
\begin{eqnarray}
\label{G1_exp} G_1(X, Y) 
 &=& \beta_e(kn_\ell) + \dfrac{1}{\pi}  \ln |X-Y| + G_H(X,Y) + r_1(X, Y), \\
\label{G2_exp} G_2(X, Y) 
 &=& \beta_e(k) + \dfrac{1}{\pi}  \ln |X-Y| + r_2(X, Y), 
\end{eqnarray}
where 
$$ \beta_e(k)=  \dfrac{1}{\pi}(\ln k + \gamma_0) + \dfrac{1}{\pi} \ln \delta, \quad r_1 = O(\delta^2|\ln\delta|), \quad r_2 = O(\delta^2|\ln\delta|). $$
In the above, $\gamma_0=c_0-\ln2 - i\pi/2$, with $c_0$ as the Euler constant.
In addition, it can be shown that (cf. Lemma 3.1 in \cite{lin_zhang17})
\begin{eqnarray}
 \label{Gi_exp} G^i(X, Y)  &=& \beta_i(kn_s) + G_0^i(X, Y)  + r_i(X, Y),   \\ 
 \label{tGi_exp} \tilde G^i(X, Y) &= &  \tilde \beta(kn_s) + \tilde r_i(X,Y), 
\end{eqnarray}
where
\begin{eqnarray*}
&& \beta_i(k)= \dfrac{\cot k }{k \delta} +  \dfrac{2\ln 2}{\pi},  \quad  \tilde \beta_i(k) = \dfrac{1}{(k\sin k) \delta}, \\
&& G_0^i(X, Y) = \dfrac{1}{\pi} \bigg[ \ln \left(\abs{\sin \left(\frac{\pi(X+Y)}{2}\right)}\right) + \ln \left(\abs{\sin \left(\frac{\pi(X-Y)}{2}\right)}\right) \bigg], \\
&& r_i = O(\delta^2), \quad \tilde r_i = O(e^{-1/\delta}).
\end{eqnarray*}

Let us introduce the following quantities of order $O(1)$:
$$ \beta_1(k) = \delta \cdot \big (\eta_1\beta_e(kn_\ell) + \beta_i(kn_s)\big), \quad  
\beta_2(k) = \delta \cdot \big (\eta_2 \beta_e(k) + \beta_i(kn_s) \big), \quad
\tilde \beta(k) = \delta  \tilde \beta_i(kn_s).  $$
For a given function $\varphi$, we define the operator $P$ such that $P \varphi(X) = \langle \varphi, 1 \rangle$.
We also introduce the integral operators $S_H$ and $S$ over the interval $(0,1)$ with the kernel 
\begin{eqnarray}
 s_H(k;X, Y)&=& \frac{\eta_1}{\pi} \ln |X-Y| + G_0^i(X,Y) + \eta_1 G_H(X,Y),  \label{eq:sH} \\
 s(X, Y)&=& \frac{\eta_2}{\pi} \ln |X-Y| + G_0^i(X,Y)  \label{eq:s}
\end{eqnarray}
respectively. 
Then by applying the expansions \eqref{G1_exp} - \eqref{tGi_exp}, the integral operator in the system \eqref{eq-scattering3} can be expanded as
\begin{equation*}
\left[
\begin{array}{cc}
\eta_1 T_1^e+T^i    & \tilde{T}^i \\
\tilde{T}^i  & \eta_2 T_2^e+T^i 
\end{array}
\right] 
= 
\frac{1}{\delta}
\left[
\begin{array}{cc}
\beta_1 P    & \tilde \beta P \\
\tilde \beta P  & \beta_2 P 
\end{array}
\right] 
 + 
 \left[
\begin{array}{cc}
 S_H  &  0 \\
 0  &  S 
\end{array}
\right] 
 +
\left[
\begin{array}{cc}
S_{H,\infty}    & \tilde S_\infty \\
\tilde S_\infty  & S_\infty
\end{array}
\right] =: \mathbb{P} + \mathbb{S}_0 + \mathbb{S}_\infty, 
\end{equation*}
where $S_\infty$, $S_{H,\infty}$ and $\tilde S_\infty$ are the integral operators with kernels given by the corresponding
high-order terms of the expansions \eqref{G1_exp} - \eqref{tGi_exp}. 
It is clear that $\norm{S_\infty} = O(\delta^2|\ln\delta|)$, $\norm{S_{H,\infty}} = O(\delta^2|\ln\delta|)$,
and $\norm{\tilde{S}_\infty} = O(e^{-1/\delta})$.
As such, the homogeneous version of \eqref{eq-scattering3}
is recast as
\begin{equation}\label{eq:hom_scattering}
(\mathbb{P} + \mathbb{S}_0 + \mathbb{S}_\infty)\boldsymbol{\varphi}=0 \quad \mbox{where} \; \boldsymbol{\varphi} = [\varphi_1,\varphi_2]^T.
\end{equation}

\subsection{Sensitivity of resonance frequency} \label{sec-resonance}
\subsubsection{Asymptotic expansion of resonances}\label{sec:resonance_exp}
We first derive the resonance frequency based on the homogeneous equation \eqref{eq:hom_scattering}.
It can be shown that $\mathbb{S}_0$ is invertible. Let $\mathbb{S}=\mathbb{S}_0 + \mathbb{S}_\infty$,
then due to the smallness of $\mathbb{S}_\infty$, the operator $\mathbb{S}$ is also invertible, and there holds $\mathbb{S}^{-1} =\mathbb{S}_0^{-1} + O(\delta^2|\ln\delta|)$.
Set $\mathbf{e}_1 = [1, 0]^T$ and $\mathbf{e}_2 = [0, 1]^T$. By applying $\delta\mathbb{S}^{-1}$ on both sides of \eqref{eq:hom_scattering}
and projecting on the subspace spanned by $\mathbf{e}_1$ and $\mathbf{e}_2$ respectively, one obtains the following system of equations:
\begin{equation*}
\mathbb{M}\left[
 \begin{array}{llll}
\langle  \varphi_1, 1 \rangle  \\
\langle  \varphi_2, 1 \rangle
\end{array}
\right] =0,
\quad \mbox{where} \;
\mathbb{M} = 
\left[ 
\begin{array}{llll}
 \langle \mathbb{S}^{-1}  \mathbf{e}_1,  \mathbf{e}_1 \rangle  &  \langle \mathbb{S}^{-1}  \mathbf{e}_2,  \mathbf{e}_1\rangle \\
 \langle \mathbb{S}^{-1}  \mathbf{e}_1,  \mathbf{e}_2 \rangle  &  \langle \mathbb{S}^{-1}  \mathbf{e}_2,  \mathbf{e}_2 \rangle 
 \end{array}
\right] 
\left[
\begin{array}{cc}
\beta_1       &   \tilde \beta  \\
\tilde \beta   &  \beta_2 
\end{array}
\right] 
+
\left[
\begin{array}{llll}
\delta & 0 \\
0 & \delta
\end{array}
\right].
\end{equation*}
Therefore, the resonances are those $k$ such that $\det\mathbb{M}(k)=0$. 

To this end,
we introduce an approximate matrix 
\begin{equation*}
\tilde{\mathbb{M}} = 
\left[ 
\begin{array}{llll}
 \alpha_H  &  0 \\
 0  &  \alpha
 \end{array}
\right] 
\left[
\begin{array}{cc}
\beta_1       &   \tilde \beta  \\
\tilde \beta   &  \beta_2 
\end{array}
\right] 
+
\left[
\begin{array}{llll}
\delta & 0 \\
0 & \delta
\end{array}
\right],
\end{equation*}
where
\begin{equation}\label{eq:alpha}
\alpha_H(k) =  \langle S_H^{-1}  1,  1 \rangle, \quad \alpha=\langle S^{-1}  1,  1 \rangle.
\end{equation}
Using the relations
\begin{eqnarray*}
&& \langle \mathbb{S}^{-1}  \mathbf{e}_1,  \mathbf{e}_1 \rangle = \alpha_H + O(\delta^2|\ln\delta|), \quad 
 \langle \mathbb{S}^{-1}  \mathbf{e}_1,  \mathbf{e}_2 \rangle = O(\delta^2|\ln\delta|), \\
&& \langle \mathbb{S}^{-1}  \mathbf{e}_2,  \mathbf{e}_1 \rangle = O(\delta^2|\ln\delta|), \quad
 \langle \mathbb{S}^{-1}  \mathbf{e}_2,  \mathbf{e}_2 \rangle = \alpha + O(\delta^2|\ln\delta|), 
\end{eqnarray*}
a sensitivity analysis leads to the relation 
\begin{equation}\label{eq:det_M}
\det\mathbb{M}= \det\tilde{\mathbb{M}} \cdot \big (1+ O(\delta^2|\ln\delta|) \big) + O(\delta^3|\ln\delta|).
\end{equation}
From Rouche's theorem, we deduce that the leading-orders of resonances are the roots of $\det\tilde{\mathbb{M}}(k) =0$.
A direct calculation yields
\begin{equation}\label{eq:det_M_hat}
\det\tilde{\mathbb{M}}(k) = \alpha_H\alpha \big(\beta_1(k)\beta_2(k) - \tilde\beta^2(k)\big) + (\alpha_H\beta_1(k) + \alpha\beta_2(k))\delta + \delta^2,
\end{equation}
which can be further decomposed as
\begin{equation}\label{eq:det_tM}
\det\tilde{\mathbb{M}}(k) = \tilde f(k) + \tilde r(k),
\end{equation}
where
\begin{eqnarray*}
\tilde f(k) &=& \left[ \alpha_H \left(\frac{\beta_1+\beta_2}{2} - \tilde\beta\right) + \delta   \right] 
   \cdot \left[ \alpha \left(\frac{\beta_1+\beta_2}{2} + \tilde\beta\right) + \delta   \right] 
   + (\alpha_H-\alpha) \tilde\beta\delta, \\
\tilde r(k) &=& -\alpha_H\alpha \left(\frac{\beta_1-\beta_2}{2}\right)^2 + \frac{\alpha_H-\alpha}{2}(\beta_1-\beta_2) \delta = O(\delta^2 \ln^2\delta). 
\end{eqnarray*}
Therefore, the leading-order of resonances can be obtained by solving for the roots of $\tilde f(k)=0$.

A close examination of $\tilde f(k)$ reveals that its roots lie close to $m\pi/n_s$ ($m=1, 2, 3, \cdots$), when
$\frac{\beta_1+\beta_2}{2} - \tilde\beta = 0$ or $\frac{\beta_1+\beta_2}{2} + \tilde\beta = 0$.
For odd $m$, let us rewrite the equation $\tilde f(k)=0$ as
$$  \alpha \left(\frac{\beta_1+\beta_2}{2} + \tilde\beta\right) + \delta 
   + \frac{(\alpha_H-\alpha) \tilde\beta}{\alpha_H \left(\frac{\beta_1+\beta_2}{2} - \tilde\beta \right) + \delta} \cdot \delta  = 0, $$ 
which takes the following explicit form:
\begin{equation}
\frac{\cos (kn_s) + 1 }{kn_s \sin(k n_s)}  + \frac{\eta_1+\eta_2}{2\pi} \cdot \delta \ln \delta + c_H(k) \cdot \delta 
+O(\delta^2) = 0,
\end{equation} 
where
\begin{equation}\label{eq:cH1}
c_H(k) = \frac{1}{2\pi} (\eta_1\ln (kn_\ell) + \eta_2 \ln(k) ) + \frac{\eta_1+\eta_2}{2\pi} \gamma_0 + \frac{2}{\pi} \ln 2 +
\dfrac{1}{\alpha}  +  \left(\frac{1}{\alpha}-\frac{1}{\alpha_H(k)}\right) \cdot \frac{1}{\cos(kn_s)-1}.
\end{equation}   
By performing the expansion of above equation at $m\pi/n_s$ and noting that the leading-order term $\frac{\cos (m\pi) + 1 }{kn_s \sin(m\pi)}=0$, 
we obtain the asymptotic expansion of its roots $k_H^{(m)}$ ($m=1, 3,  5 \cdots$)
for $m\delta \ll 1$:
\begin{equation}\label{eq:km_exp1}
k_H^{(m)} \cdot n_s =  m \pi + m (\eta_1+\eta_2) \cdot \delta \ln\delta + 
2m\pi c_H(m \pi/n_s) \cdot \delta  + O(\delta^2\ln^2\delta).
\end{equation}
In view of the relations \eqref{eq:det_M} and \eqref{eq:det_tM}, the above expansion also holds for the roots of $\det\mathbb{M}(k)=0$,
or equivalently, the resonances of the scattering problem.

\begin{figure}[!htbp]
\begin{center}
\includegraphics[height=5cm]{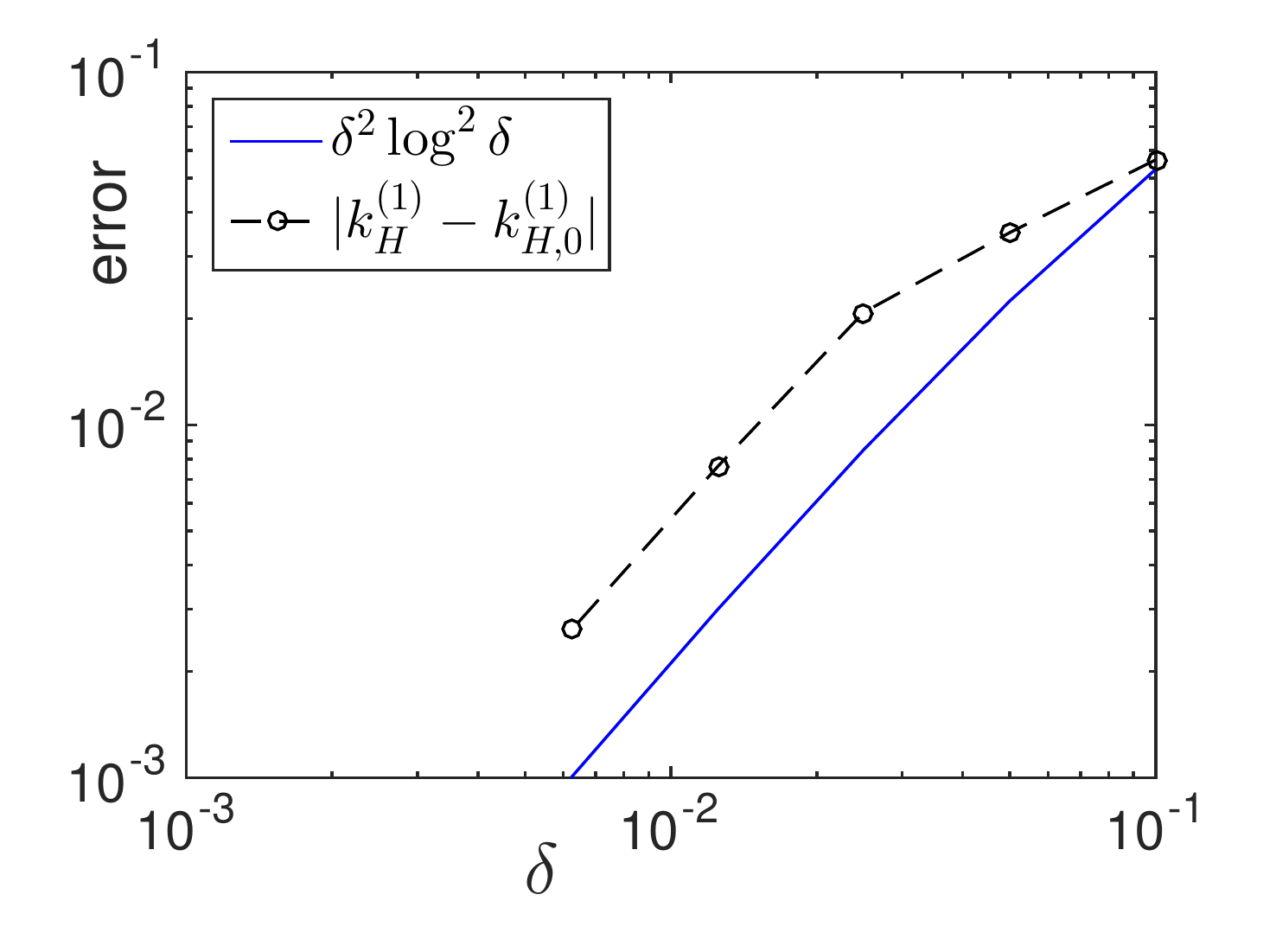} 
\includegraphics[height=5cm]{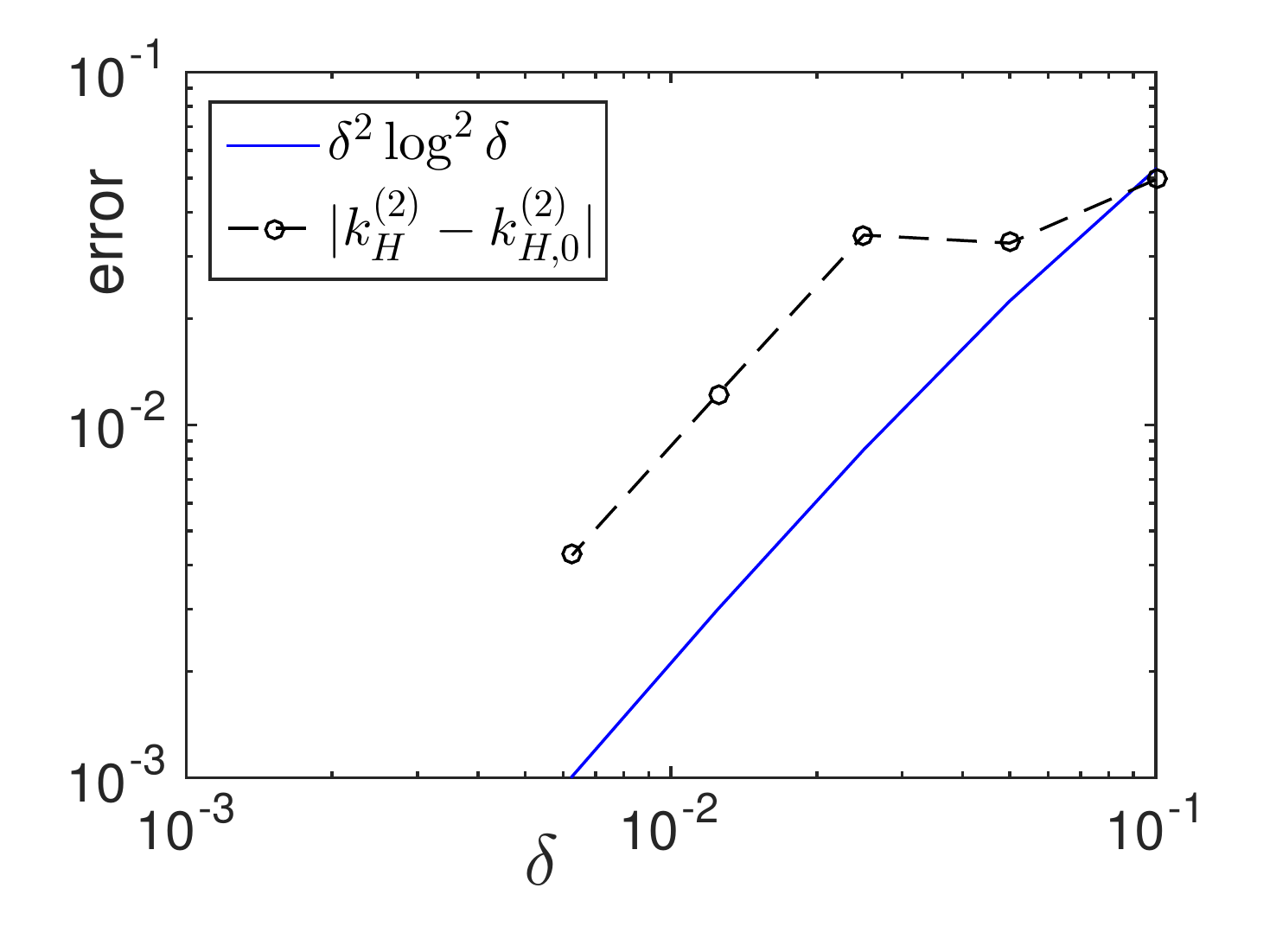}
\caption{Accuracy of the asymptotic expansion formulas for the resonances $k_H^{(1)}$ and $k_H^{(2)}$.
The dash lines represent the error
$|k_H^{(1)}-k_{H,0}^{(1)}|$ and $|k_H^{(2)}-k_{H,0}^{(2)}|$ respectively, in which $k_{H,0}^{(1)}$ and $k_{H,0}^{(2)}$ are the values obtained from the asymptotic formulas
with the high-order terms $O(\delta^2\ln^2\delta)$ being dropped. 
We set $\varepsilon_\ell=2$ and $\varepsilon_s=1$ in the calculations, and $k_H^{(1)}$ and $k_H^{(2)}$ are obained numerically with high-order accuray. }\label{fig:error_asy_single_slit}
\end{center}
\end{figure}

Similarly, for even $m$,  by rewriting $\tilde f(k)=0$ as
$$  \alpha_H \left(\frac{\beta_1+\beta_2}{2}  - \tilde\beta\right) + \delta 
   + \frac{(\alpha_H-\alpha) \tilde\beta}{\alpha \left(\frac{\beta_1+\beta_2}{2} + \tilde\beta \right) + \delta} \cdot \delta  = 0, $$ 
the Taylor expansion at $m\pi/n_s$ leads to the expansion of its roots $k_m$ ($m=2, 4,  6 \cdots$):
\begin{equation}\label{eq:km_exp2}
k_H^{(m)}\cdot n_s =  m \pi + m (\eta_1+\eta_2) \cdot \delta \ln\delta + 2m\pi c_H(m \pi/n_s) \cdot \delta  + O(\delta^2\ln^2\delta),
\end{equation}
where 
\begin{equation}\label{eq:cH2}
c_H(k) = \frac{1}{2\pi} (\eta_1\ln (kn_\ell) + \eta_2 \ln(k) ) + \frac{\eta_1+\eta_2}{2\pi} \gamma_0 + \frac{2}{\pi} \ln 2 +
\dfrac{1}{\alpha_H(k)}  +  \left(\frac{1}{\alpha}-\frac{1}{\alpha_H(k)}\right) \cdot \frac{1}{\cos(kn_s)+1}.
\end{equation}
Thus \eqref{eq:km_exp2} gives the expansion of resonances for even $m$. Figure \ref{fig:error_asy_single_slit} demonstrates the accuracy
of the asympototic expansions \eqref{eq:km_exp1} and \eqref{eq:km_exp2} for the resonance frequencies. The plot confirms the order 
$O(\delta^2\ln^2\delta)$ obtained in the asymptotic expansions.

\subsubsection{Sensitivity analysis of resonance frequency}\label{sec:res_sensitivity}
The sensitivity of the resonance $k_H^{(m)}$ with respect to the thickness of the thin layer is defined as $ \frac{\partial k_H^{(m)}}{\partial H} = \lim_{\Delta H \to 0} \frac{k_{H+\Delta H}^{(m)} - k_H^{(m)}}{\Delta H}$.
From the expansion of  $k_H^{(m)}$ in \eqref{eq:km_exp1} and \eqref{eq:km_exp2}, this boils down to the sensitivity analysis of the coefficient 
$\alpha_H$, namely $ \frac{\partial \alpha_H}{\partial H}$.   Recall that $\alpha_H =  \langle S_H^{-1}  1,  1 \rangle$, and the kernel of the operator
$S_H$ is given in \eqref{eq:sH},
thus we have
$$ \frac{\partial \alpha_H}{\partial H} =  \left\langle \frac{ \partial S_H^{-1}}{\partial H} 1, 1 \right\rangle =  \left\langle S_H^{-1}\frac{ \partial S_H}{\partial H}S_H^{-1} 1 , 1 \right\rangle 
= \left\langle  \frac{ \partial S_H}{\partial H} \varphi_H, \varphi_H \right\rangle, 
$$
where $\varphi_H$ is the solution for the integral equation $S_H\varphi_H = 1$.
In view of the expression \eqref{eq:GH} for $G_H(X,Y)$, the kernel of $\frac{ \partial S_H}{\partial H}$ is given explicitly by
\begin{equation} \label{eq111}  
\frac{ \partial G_H}{\partial H} = \frac{4}{\pi}\int_0^\infty 
\frac{ A_{+}A_{-} e^{i2\rho(kn_\ell,\xi)H} \cdot 
\cos(\delta\xi(X-Y))}{ \big(A_+ - A_- e^{i2\rho(kn_\ell,\xi)H}\big)^2 } d \xi,
\end{equation}
where
$A_{+}= A_{+}(k, \xi), A_{-}= A_{-}(k, \xi)$ and $\rho(kn_\ell,\xi)$ are defined in (\ref{Apm}).

To compute $\frac{\partial \alpha_H}{\partial H}$, we divide the whole Sommerfeld integral frequency band as 
$$\Lambda_1 = \{\xi; \, 0\le\xi \le kn_\ell+1\} \quad \mbox{and} \quad \Lambda_2 = \{\xi; \, \xi > kn_\ell+1 \},$$
and decompose
$\frac{ \partial G_H}{\partial H}$ as $\frac{ \partial G_H}{\partial H}=\frac{ \partial G_{1,H}}{\partial H}+\frac{ \partial G_{2,H}}{\partial H}$, where
\begin{eqnarray*}
\frac{ \partial G_{j,H}}{\partial H} = \frac{4}{\pi} \int_{\Lambda_j} 
\frac{ A_{+}A_{-} e^{i2\rho(kn_\ell,\xi)H} \cdot 
\cos(\delta\xi(X-Y))}{ \big(A_+ - A_- e^{i2\rho(kn_\ell,\xi)H}\big)^2 } d \xi, \,  \quad j=1, 2.
\end{eqnarray*}
Correspondingly,  $\frac{\partial \alpha_H}{\partial H}$ is decomposed into
$$
\frac{\partial \alpha_{1, H}}{\partial H} =
 \left\langle  \frac{ \partial S_{1,H}}{\partial H} \varphi_H, \varphi_H \right\rangle   
\quad \mbox{and} \quad \frac{\partial \alpha_{2, H}}{\partial H} 
= \left\langle  \frac{ \partial S_{2,H}}{\partial H} \varphi_H, \varphi_H \right\rangle, 
$$
in which $ \frac{ \partial S_{j,H}}{\partial H}$ is the integral operator with the kernel $\frac{ \partial G_{j,H}}{\partial H}$. 
Since $\Lambda_1$ has finite bandwidth,
it is clear that $\frac{\partial G_{1, H}}{\partial H} = O(1)$ and hence $$\frac{\partial \alpha_{1, H}}{\partial H} = O(1).$$
We now compute  $\frac{\partial \alpha_{2, H}}{\partial H}$ by investigating the kernel of $\frac{\partial S_{2, H}}{\partial H}$, which
consists of all evanescent wave modes with the momentum $\xi$
in the frequency band  $\Lambda_{2}$.
Note that for $\xi\in\Lambda_{2}$, there hold
\begin{eqnarray*}
&& A_+ = i \xi \left(\sqrt{1- \frac{k^2n_l^2}{\xi^2}} + \sqrt{1- \frac{k^2}{\xi^2}}\varepsilon_\ell \right), \\ 
&& A_- = i \xi \left(\sqrt{1- \frac{k^2n_l^2}{\xi^2}} - \sqrt{1- \frac{k^2}{\xi^2}}\varepsilon_\ell \right), \\
&& e^{i2\rho(kn_\ell,\xi)H}  = e^{-2\xi H \sqrt{1- \frac{k^2n_l^2}{\xi^2} }}.
\end{eqnarray*}
Let 
$$
F(\xi) = \frac{ A_{+}A_{-}}{ \big(A_+ - A_- e^{i2\rho(kn_\ell,\xi)H}\big)^2 } b(\xi),
$$
where
$$
b(\xi)= e^{2\xi H - 2 \xi H \sqrt{1- \frac{k^2n_l^2}{\xi^2}}} = 
e^{\frac{2k^2 n_l^2H}{\xi \big(\sqrt{1- k^2n_l^2/\xi^2}+1 \big)}}.
$$
Then 
\begin{equation}\label{eq:int_dG2dH}
\frac{ \partial G_{2,H}}{\partial H}  =  \int_{\xi > kn_l +1} F(\xi) e^{-2\xi H}\cdot 
\cos(\delta\xi(X-Y)) d \xi = \Re \int_{\xi > kn_l +1} F(\xi) e^{-\xi (2H +  i \delta (X-Y))} d \xi. 
\end{equation}
Using the expansions
$$
A_+ = i \xi \left(1+ \varepsilon_l + O(\frac{1}{\xi^2})\right), \quad 
A_- = i \xi \left(1- \varepsilon_l + O(\frac{1}{\xi^2})\right), \quad 
b(\xi) = 1 + O\left(\frac{1}{\xi}\right),
$$
we can decompose $F(\xi)$ as
$$
F(\xi) =  F_0(\xi) + F_1(\xi)=: \frac{a}{(1- a e^{-2\xi H})^2} + O\left(\frac{1}{\xi}\right),
$$
where 
$$
a = \frac{1-\varepsilon_l }{1+ \varepsilon_l}.
$$

We now estimate $\int_{\xi > kn_l +1} F_j(\xi) e^{-\xi (2H +  i \delta (X-Y))} d \xi$ for $j=0, 1$. 
Since 
$$
\int_{\xi > kn_l +1} \frac{1}{\xi} e^{-2H \xi} d\xi = O(\ln H),
$$
it follows that
\begin{equation}\label{eq:int_F1}
\int_{\xi > kn_l +1} F_1(\xi) e^{-\xi (2H +  i \delta (X-Y))} d \xi = O(\ln H).
\end{equation}
On the other hand, note that
$$
\frac{1}{(1- a e^{-2\xi H})^2} = \sum_{n=0}^{\infty} (n+1) a^n e^{-2n\xi H}, 
$$
we obtain
\begin{eqnarray}\label{eq:int_F0}
\int_{\xi > kn_l +1} F_0(\xi) e^{-\xi (2H +  i \delta (X-Y))} d \xi &=&a \sum_{n=0}^{\infty} \int_{\xi > kn_l +1} (n+1) a^n e^{-2n\xi H}  e^{-\xi (2H +  i \delta (X-Y))} d \xi  \nonumber \\
&=&a \sum_{n=0}^{\infty} (n+1)a^n \frac{e^{(-2(n+1)H -i \delta (X-Y))(kn_l+1)}}{2(n+1)H + i \delta (X-Y)}  \nonumber \\
& =&  O\left(\frac{1}{\sqrt{H^2 + \delta^2 (X-Y)^2 }}\right). \label{eq:int_F0}
\end{eqnarray}

By substituting \eqref{eq:int_F1} and \eqref{eq:int_F0} into \eqref{eq:int_dG2dH}, we obtain the sensivitiy for $\frac{ \partial G_{2, H}}{\partial H} $:
\begin{equation*} 
\frac{ \partial G_{2, H}}{\partial H} =  O(\ln H) + O\left(\frac{1}{\sqrt{H^2 + \delta^2 (X-Y)^2 }}\right).
\end{equation*}
It follows that for $H > \delta$, 
\begin{equation} \label{g-2-est1}
\frac{ \partial G_{2, H}}{\partial H} = O(\ln H) + O (1/H)= O (1/H),
\end{equation}
and for $0< H \leq \delta$, 
\begin{equation} \label{g-2-est2}
\frac{ \partial G_{2, H}}{\partial H} = O(\ln H) + O \left(\frac{1}{\delta} \frac{1}{\sqrt{(H/\delta)^2 + (X-Y)^2 }} \right).
\end{equation}

We now estimate 
$$\frac{\partial \alpha_{2, H}}{\partial H} 
= \left\langle  \frac{ \partial S_{2,H}}{\partial H} \varphi_H, \varphi_H \right\rangle = \displaystyle{\int_0^1 \int_0^1 \frac{ \partial G_{2, H}}{\partial H}(X, Y) \varphi_H(X)\varphi_H(Y)dX dY}.$$
Since $H \ll 1$, it follows that
\begin{equation}\label{eq:dalpha2_dH}
\frac{\partial \alpha_{2, H}}{\partial H} \approx \int_0^1 \int_0^1 \frac{ \partial G_{2, H}}{\partial H}(X, Y) \varphi_0(X)\varphi_0(Y)dX dY,
\end{equation}
where $\varphi_0$ is the unique solution to the integral equation $S_0 \varphi_0 =1$. It is known that $\varphi_0$ is a smooth function over the interval $(0, 1)$
, and it attains the singularity of the type $\frac{1}{\sqrt{X}}$ and $\frac{1}{\sqrt{1-X}}$ near $X=0$ and $X=1$ respectively (see, for instance, \cite{Costabel-2007}). Hence we may write $\varphi_0$ as
\begin{equation} \label{eq:phi0}
\varphi_0(X) = O \left(\frac{1}{\sqrt{X}}\right) +  O \left(\frac{1}{\sqrt{1-X}}\right).
\end{equation}
Inserting \eqref{g-2-est1} and \eqref{eq:phi0} into \eqref{eq:dalpha2_dH}, we obtain that for $H > \delta$,
$$
\frac{\partial \alpha_{2, H}}{\partial H} = O\left(\frac{1}{H}\right). 
$$
On the other hand, for $0< H \leq  \delta$, by substituting \eqref{g-2-est2} and \eqref{eq:phi0} into \eqref{eq:dalpha2_dH} and using the estimate
$$
\int_0^1 \int_0^1 \frac{1}{\sqrt{t^2 + (X-Y)^2 }}  \frac{1}{\sqrt{X}}\frac{1}{\sqrt{Y}}dX dY = O(\ln t +1), \quad 0<t\le1,
$$
we obtain
$$
\frac{\partial \alpha_{2, H}}{\partial H} =  O(\ln H) + \frac{1}{\delta} \cdot O\left( 1 + \ln (H/\delta) \right) =  \frac{1}{\delta} \cdot O\left( 1 + \ln (H/\delta) \right),
$$

\begin{figure}[!htbp]
\begin{center}
\includegraphics[height=4.5cm]{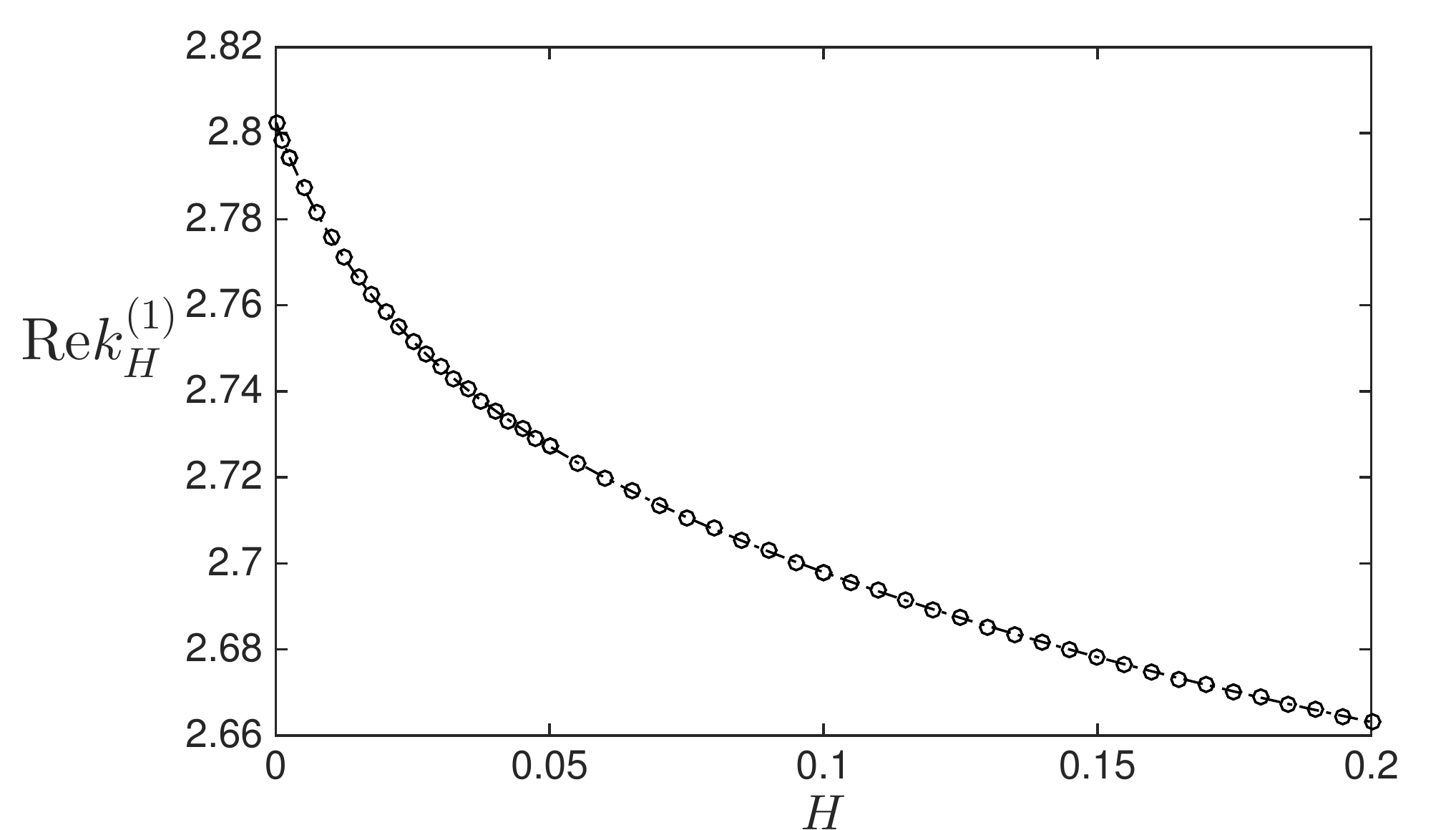} 
\includegraphics[height=4.5cm]{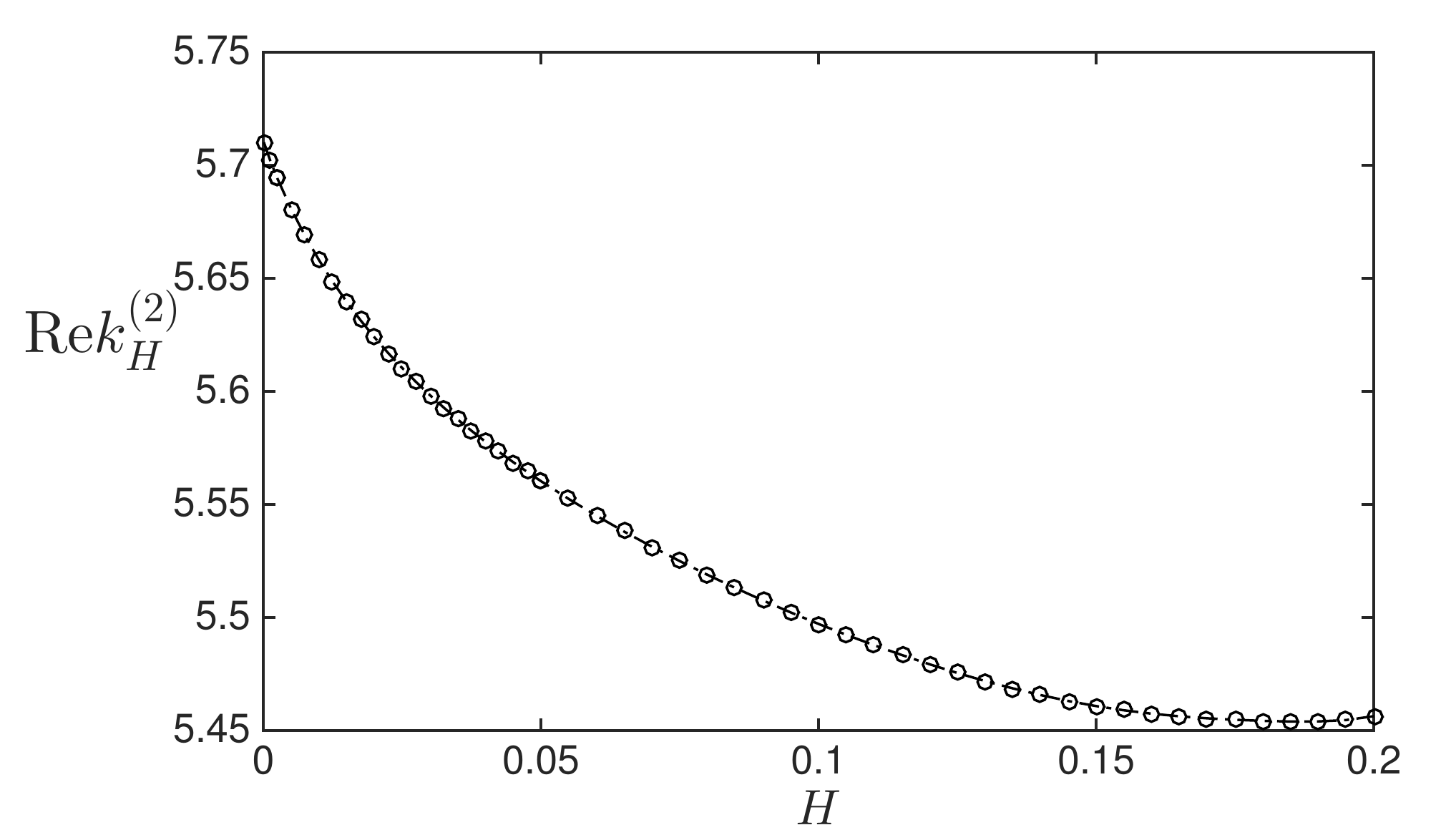}
\caption{The resonance frequencies $\Re k_H^{(1)}$ and $\Re k_H^{(2)}$ for various thickness $H\in[10^{-5}, 0.2]$ when $\delta=0.05$.
Here $\varepsilon_\ell=2$ and $\varepsilon_s=1$. }\label{fig:kh_single_slit}
\end{center}
\end{figure}

Finally, by inserting the sensitivity of $\alpha_H = \alpha_{1, H} + \alpha_{2, H}$
into the expansions in \eqref{eq:km_exp1} and \eqref{eq:km_exp2} for the resonances, it follows that the sensitivity of the resonance frequency is given by 
$ \frac{\partial k_H^{(m)}}{\partial H} = O(\delta/H)$ if $H > \delta$ and $ \frac{\partial k_H^{(m)}}{\partial H} = O(1+\ln (H/\delta))$  if $0<H < \delta$. 
We observe that the spectral sensitivity decreases if $H$ increases. 
This is illustrated in Figure \ref{fig:kh_single_slit} where $k_H^{(1)}$ and  $k_H^{(2)}$ are depicted for various thickness $H\in[10^{-5}, 0.2]$ when $\delta=0.05$. 

\section{Sensitivity of resonance frequency for the periodic nano-slit structure}
Denote the period of the metallic structure by $d$ and the reciprocal lattice constant by $b=2\pi/d$.
Let $\kappa=k\sin\theta$ be the Bloch wavenumber. 
Due to the periodicity of the structure, we impose that the solution $u$ is the quasi-periodic such that
$u(x_1+d,x_2)$ = $e^{i\kappa d}u(x_1,x_2)$.
By applying the Green's formula in the $\Omega_H$ and $\Omega_2$ using the quasi-period Green's functions
with the momentum $\kappa$ and taking the limit to the slit apertures,
one obtains a similar integral formulation to the single nano-slit case:
\begin{equation}\label{eq-scattering_periodic}
\left[
\begin{array}{cc}
\eta_1 T_1^p+T^i    & \tilde{T}^i \\
\tilde{T}^i  & \eta_2 T_2^p+T^i 
\end{array}
\right] \left[
\begin{array}{llll}
\varphi_1    \\
\varphi_2 
\end{array}
\right]=\left[
\begin{array}{llll}
0  \\
f/\delta  
\end{array}
\right],
\end{equation}
where $T_1^p$ and $T_2^p$, $T^i$ are integral operators defined for $X\in (0,1)$ with quasi-periodic kernels $G_1^p(X, Y)$ and $G_2^p(X, Y)$.
Let $\kappa_n=\kappa+\dfrac{2\pi n}{d}$, then the explicit expression of quasi-periodic Green's functions $G_1^p(X, Y)$ and $G_2^p(X, Y)$
are given by
\begin{eqnarray*}
G_1^p(X, Y) &=& -\dfrac{i}{d} \sum_{n=-\infty}^{\infty} \dfrac{1}{\rho(kn_\ell,\kappa_n)   } e^{i \kappa_n\delta(X-Y)} + G_H^p(k;X,Y) \quad \; \mbox{for} \; X, Y \in
\Gamma_{1,\delta},  \\
G_2^p(X, Y) &=& -\dfrac{i}{d} \sum_{n=-\infty}^{\infty} \dfrac{1}{\rho(k,\kappa_n)   } e^{i \kappa_n\delta(X-Y)} \quad \; \mbox{for} \; X, Y \in
\Gamma_{2,\delta},
\end{eqnarray*}
in which
\begin{equation}\label{eq:GHp}
G_H^p(k;X,Y)= -\frac{2i}{d}  \sum_{n=-\infty}^\infty \dfrac{  A_-(k,\kappa_n)  e^{i2\rho(kn_\ell,\kappa_n) H }  e^{i\kappa_n\delta(X-Y)} }
{\rho(kn_\ell,\kappa_n)  \big( A_+(k,\kappa_n) - A_-(k,\kappa_n) e^{i2\rho(kn_\ell,\kappa_n) H } \big) },
\end{equation}
and $A_\pm$ and $\rho$ are defined in \eqref{Apm}.

If $\delta\ll1$ and $k$ is away from the Rayleigh anomaly frequencies satisfying $k=\kappa_n$ or $kn_\ell=\kappa_n$ for certain $n$, 
then it can be shown that the Green's functions admit the following asymptotic expansions (cf. Lemma 3.1 in \cite{lin_zhang18_1}):
\begin{eqnarray}
\label{G1p_exp} G_1^p(X, Y) 
 &=& \beta_{e}^p(kn_\ell,\kappa) + \dfrac{1}{\pi}  \ln |X-Y| + G_H^p(k;X,Y) + r_1(X, Y), \\
\label{G2p_exp} G_2^p(X, Y) 
 &=& \beta_{e}^p(k,\kappa) + \dfrac{1}{\pi}  \ln |X-Y| + r_2(X, Y), 
\end{eqnarray}
where
$$ \beta_{e}^p(k,\kappa)=\dfrac{1}{\pi} \left(\ln\delta + \ln 2 + \ln\dfrac{\pi}{d}\right) - \dfrac{i}{d} \dfrac{1}{\rho(k,\kappa)} + 
\sum_{n\neq 0} \left(\dfrac{1}{2\pi}\dfrac{1}{|n|} - \dfrac{i}{d}  \dfrac{1}{\rho(k,\kappa_n)}\right), $$
and $r_1 = O(\delta)$, $r_2 = O(\delta)$. By decomposing the integral operators $T_1^p$, $T_2^p$, and $T^i$ according to the expansions
of their kernels $G_1^p$, $G_2^p$ and $G^i$, a decomposition parallel to \eqref{eq:hom_scattering} of the single slit case can be obtained for the homogeneous version of integral formulation \eqref{eq-scattering_periodic}. 
Note that for the periodic case, the operator $\mathbb{S}_0$ takes the form of 
\begin{equation*}
\mathbb{S}_0 = \left[
\begin{array}{cc}
 S_H^p  &  0 \\
 0  &  S 
\end{array}
\right],
\end{equation*}
where the integral operator $S$ remains the same, while the integral operator $S_H^p$ attains the kernel 
$$s_H^p(k;X,Y)= \frac{\eta_1}{\pi} \ln |X-Y| + G_0^i(X,Y) + \eta_1 G_H^p(k;X,Y).$$ 
Therefore, by projecting on the subspace spanned by $\mathbf{e}_1$ and $\mathbf{e}_2$ as the single nano-slit configuration, 
the resonances reduce to the roots of certain nonlinear function given by the determinant of a $2 \times 2$ matrix $\mathbb{M}^p(k,\kappa)$.

To solve for $\det\mathbb{M}^p(k,\kappa)=0$,  we define
$$\alpha_H^p =  \langle (S_H^p)^{-1}  1,  1 \rangle, \quad \alpha =  \langle S^{-1}  1,  1 \rangle, $$
and
$$ \beta_1^p(k,\kappa) = \delta \cdot \big (\eta_1\beta_{e}^p(kn_\ell,\kappa) + \beta_i(kn_s)\big), \quad  
\beta_2^p(k) = \delta \cdot \big (\eta_2 \beta_{e}^p(k,\kappa) + \beta_i(kn_s) \big). $$
Then following the same calcuations as the single slit case, the leading-order of roots near to $m\pi/n_s$ ($m=1, 3,  5 \cdots$) satisfy 
$$  \alpha \left(\frac{\beta_1^p+\beta_2^p}{2} + \tilde\beta\right) + \delta 
   + \frac{(\alpha_H^p-\alpha) \tilde\beta}{\alpha_H^p \left(\frac{\beta_1+\beta_2}{2} - \tilde\beta \right) + \delta} \cdot \delta  = 0. $$ 
From the above equation,  we obtain the asymptotic expansion of the roots $k_H^{(m)}(\kappa)$ for each $\kappa$ when $m$ is odd:
\begin{equation}\label{eq:kmp_exp1}
k_H^{(m)}(\kappa) \cdot n_s =  m \pi + m (\eta_1+\eta_2) \cdot \delta \ln\delta + 
2m\pi c_H^p(m \pi/n_s,\kappa) \cdot \delta  + O(\delta^2\ln^2\delta),
\end{equation}
where
\begin{eqnarray*}
c_H^p(k,\kappa) &=& \frac{1}{2\pi} \left((\eta_1+\eta_2+4) \ln 2 + (\eta_1+\eta_2) \ln \frac{\pi}{d} \right) 
+ \frac{\eta_1}{2}\gamma(kn_\ell,\kappa) + \frac{\eta_2}{2}\gamma(k,\kappa) \\
&& + \dfrac{1}{\alpha}  +  \left(\frac{1}{\alpha}-\frac{1}{\alpha_H^p}\right) \cdot \frac{1}{\cos(kn_s)-1}, \\
\gamma(k,\kappa) &=& - \dfrac{i}{d} \dfrac{1}{\rho(k,\kappa)} + 
\sum_{n\neq 0} \left(\dfrac{1}{2\pi}\dfrac{1}{|n|} - \dfrac{i}{d}  \dfrac{1}{\rho(k,\kappa_n)}\right).
\end{eqnarray*}
For even $m$ such that $m\delta\ll 1$, it can be obtained that 
\begin{equation}\label{eq:kmp_exp2}
k_H^{(m)}(\kappa) \cdot n_s =  m \pi + m (\eta_1+\eta_2) \cdot \delta \ln\delta + 
2m\pi c_H^p(m \pi/n_s,\kappa) \cdot \delta  + O(\delta^2\ln^2\delta),
\end{equation}
where
\begin{eqnarray*}
c_H^p(k,\kappa) &=& \frac{1}{2\pi} \left((\eta_1+\eta_2+4) \ln 2 + (\eta_1+\eta_2) \ln \frac{\pi}{d} \right) 
+ \frac{\eta_1}{2}\gamma(kn_\ell,\kappa) + \frac{\eta_2}{2}\gamma(k,\kappa) \\
&& + \dfrac{1}{\alpha_H^p}  +  \left(\frac{1}{\alpha}-\frac{1}{\alpha_H^p}\right) \cdot \frac{1}{\cos(kn_s)+1}.
\end{eqnarray*}

\begin{figure}[!htbp]
\begin{center}
\includegraphics[height=5cm]{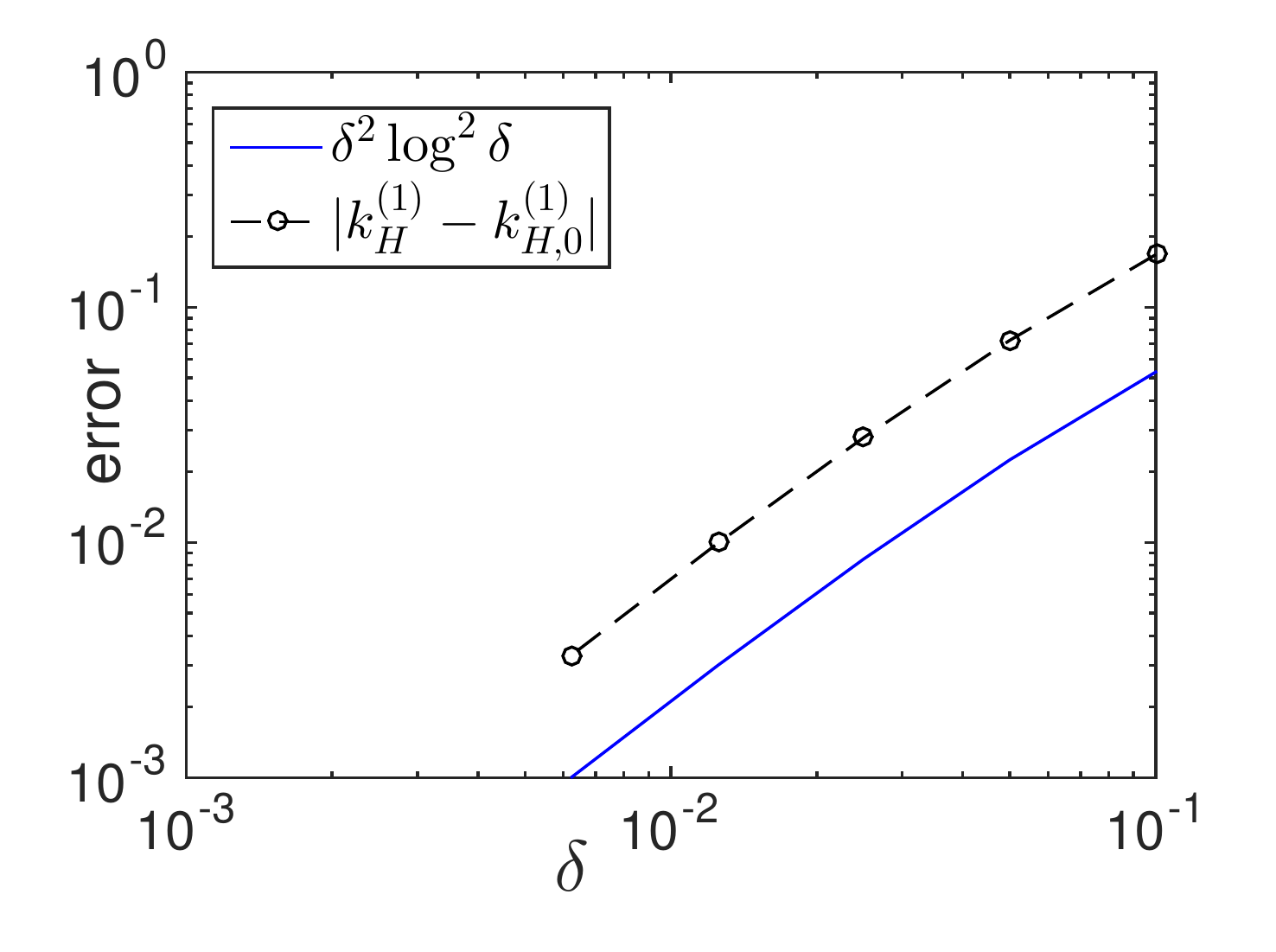} 
\includegraphics[height=5cm]{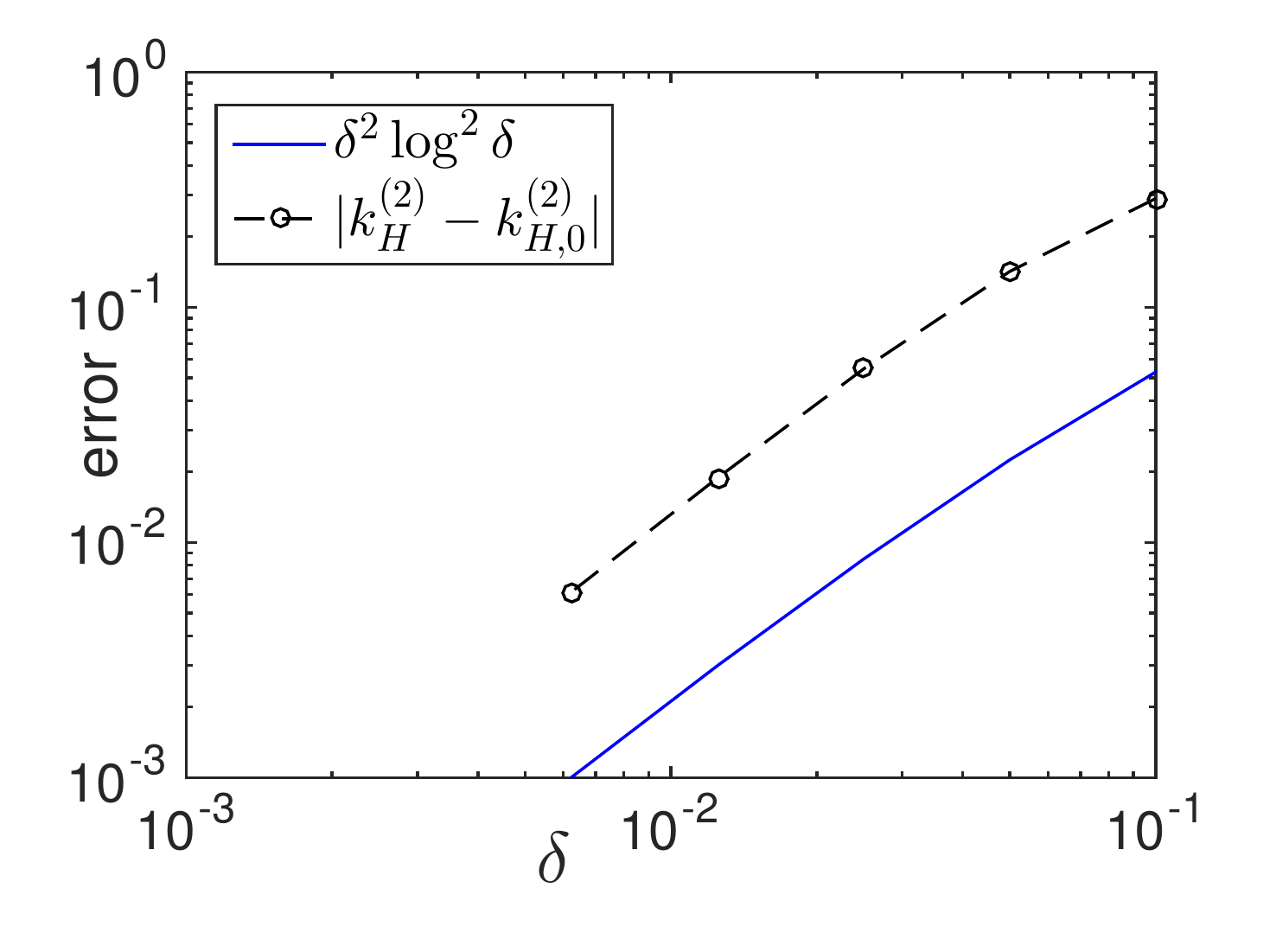}
\caption{Accuracy of the asymptotic expansion formulas for the resonances $k_H^{(1)}$ and $k_H^{(2)}$.
The dash lines represent the error
$|k_H^{(1)}-k_{H,0}^{(1)}|$ and $|k_H^{(2)}-k_{H,0}^{(2)}|$ respectively, in which $k_{H,0}^{(1)}$ and $k_{H,0}^{(2)}$ are the values obtained from the asymptotic formulas
with the high-order terms $O(\delta^2\ln^2\delta)$ being dropped. 
The period $d=1.2$, the Bloch wavenumber $\kappa=0$, and the permittivity values $\varepsilon_\ell=2$, $\varepsilon_s=1$. }\label{fig:error_asy_periodic}
\end{center}
\end{figure}

In view of the expansions \eqref{eq:kmp_exp1} and \eqref{eq:kmp_exp2} for the resonances,
the sensitivity of $k_H^{(m)}$ now reduces to the sensitivity analysis of the coefficient 
$\alpha_H^p$, namely $\frac{\partial \alpha_H^p}{\partial H}$. This boils down to the following inner product:
$$ \frac{\partial \alpha_H^p}{\partial H} =  \left\langle \frac{ \partial (S_H^p)^{-1}}{\partial H} 1, 1 \right\rangle 
=  \left\langle (S_H^p)^{-1}\frac{ \partial S_H^p}{\partial H}(S_H^p)^{-1} 1 , 1 \right\rangle 
= \left\langle  \frac{ \partial S_H^p}{\partial H} \varphi_H^p, \varphi_H^p \right\rangle,   
$$
where $\varphi_H^p$ is the solution for the integral equation $S_H^p\varphi^p = 1$.
The derivation can be proceeded by following the lines of the single slit configuration.

First, using the expression \eqref{eq:GHp}, we obtain the kernel of $\frac{ \partial S_H^p}{\partial H}$:
\begin{equation} \label{eq112} 
\frac{ \partial G_H^p}{\partial H} = \frac{4}{d}\sum_{n=-\infty}^\infty 
\frac{ A_{+}A_{-} e^{i2\rho(kn_\ell,\kappa_n)H} \cdot 
e^{i\kappa_n\delta(X-Y)}}{ \big(A_+ - A_- e^{i2\rho(kn_\ell,\kappa_n)H}\big)^2 }.
\end{equation}
where
$A_{+}= A_{+}(k, \xi), A_{-}= A_{-}(k, \xi)$ and $\rho(kn_\ell,\xi)$ are defined in (\ref{Apm}).
Let $N$ be the smallest integer satisfying $Nb>kn_\ell+|\kappa|$. 
We decompose 
$\frac{ \partial G_H^p}{\partial H}$ as $\frac{ \partial G_H^p}{\partial H}=\frac{ \partial G_{1,H}^p}{\partial H}+\frac{ \partial G_{2,H}^p}{\partial H}$, where
\begin{eqnarray*}
\frac{ \partial G_{1,H}^p}{\partial H} &=& 
\frac{4}{d}\sum_{|n|\leq N} 
\frac{ A_{+}A_{-} e^{i2\rho(kn_\ell,\kappa_n)H} \cdot e^{i\kappa_n\delta(X-Y)}}{ \big(A_+ - A_- e^{i2\rho(kn_\ell,\kappa_n)H}\big)^2 }, \\
\frac{ \partial G_{2,H}^p}{\partial H} &=& 
\frac{4}{d}\sum_{|n|> N} 
\frac{ A_{+}A_{-} e^{i2\rho(kn_\ell,\kappa_n)H} \cdot e^{i\kappa_n\delta(X-Y)} }
{ \big(A_+ - A_- e^{i2\rho(kn_\ell,\kappa_n)H}\big)^2 }. 
\end{eqnarray*}
It is clear that $\frac{ \partial G_{1,H}^p}{\partial H}$ consists of finitely many propagating modes, 
while $\frac{ \partial G_{2,H}^p}{\partial H}$ consists of infinitely many evanescent modes,
which will be dominant in the resonance sensitivity analysis.
Correspondingly,  $\frac{\partial \alpha_H}{\partial H}$ is decomposed into
$$
\frac{\partial \alpha_{1, H}}{\partial H} =
 \left\langle  \frac{ \partial S_{1,H}}{\partial H} \varphi_H, \varphi_H \right\rangle   
\quad \mbox{and} \quad \frac{\partial \alpha_{2, H}}{\partial H} 
= \left\langle  \frac{ \partial S_{2,H}}{\partial H} \varphi_H, \varphi_H \right\rangle, 
$$
in which $ \frac{ \partial S_{j,H}}{\partial H}$ is the integral operator with the kernel $\frac{ \partial G_{j,H}}{\partial H}$.
A parallel argument as in Section \ref{sec:res_sensitivity} by replacing the integration over the frequency band $\Lambda_1$ and $\Lambda_2$ 
with the sum of the series correspondingly shows that
\begin{align*}
\frac{ \partial G_{1,H}^p}{\partial H}&= O(1), \\
\frac{ \partial G_{2, H}}{\partial H} &=  O(\ln H) + O\left(\frac{1}{\sqrt{H^2 + \delta^2 (X-Y)^2 }}\right).
\end{align*}
As such we obtain the sensitivity for $\alpha_{1,H}^p$ and $\alpha_{2,H}^p$ as follows:
\begin{align*}
\frac{\partial \alpha_{1, H}^p}{\partial H} &= O(1), \\
\frac{\partial \alpha_{2, H}^p}{\partial H} &= O(1/H), \quad H > \delta, \\
\frac{\partial \alpha_{2, H}^p}{\partial H} &=   \frac{1}{\delta} \cdot O\left( 1 + \ln (H/\delta) \right), \quad 0< H \le \delta.
\end{align*}

\begin{figure}[!htbp]
\begin{center}
\includegraphics[height=4.5cm]{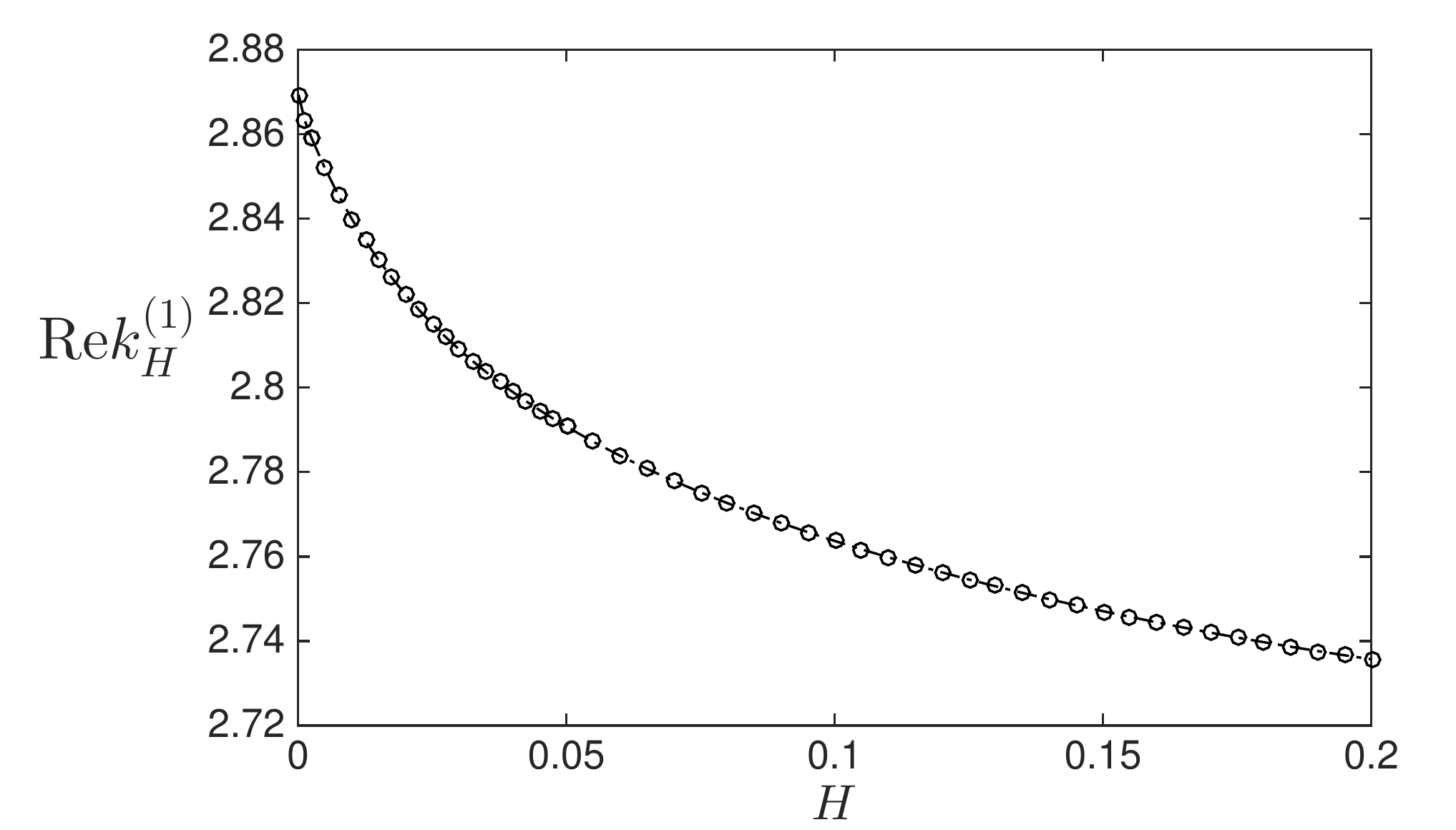} 
\includegraphics[height=4.5cm]{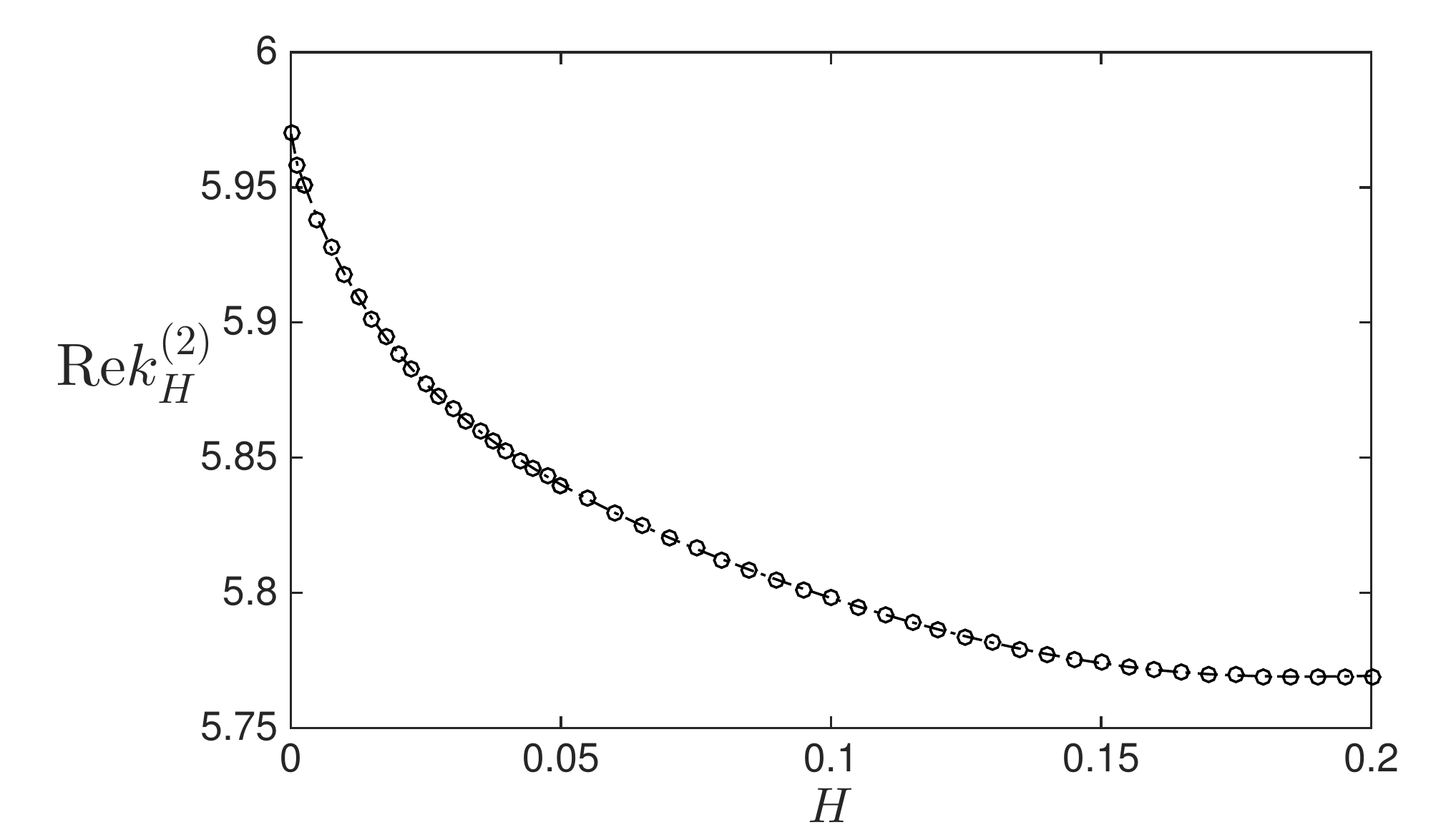}
\caption{The resonance frequencies $\Re k_H^{(1)}$ and $\Re k_H^{(2)}$ for various thickness $H\in[10^{-5}, 0.2]$ when $\delta=0.05$.
The period $d=1.2$, the Bloch wavenumber $\kappa=0$, and the permittivity values $\varepsilon_\ell=2$, $\varepsilon_s=1$. }\label{fig:kh_periodic_slits}
\end{center}
\end{figure}

By inserting the sensitivity of $\alpha_H^p = \alpha_{1,H}^p+\alpha_{2,H}^p$ into the resonance expansions \eqref{eq:kmp_exp1} and \eqref{eq:kmp_exp2}, 
it follows that for each $\kappa$, the sensitivity of the resonance frequency is given by 
$\frac{\partial k_H^{(m)}(\kappa)}{\partial H} = O(\delta/H)$ if $ H > \delta $ and $\frac{\partial k_H^{(m)}(\kappa)}{\partial H}= O(1+\ln (H/\delta))$  if $0<H < \delta$. This is similar to the sensitivity of the single slit case
studied in Section \ref{sec-resonance}, where the spectral sensitivity is reduced when $H$ increases.
Figure \ref{fig:kh_periodic_slits} plots the first two resonance frequencies for various thickness 
$H\in[10^{-5}, 0.2]$, which confirm such an assertion.

\begin{thebibliography}{99}





\bibitem{Ammari-Yu-2018}
H. Ammari, D. Choi and S. Yu,
{ \em A mathematical and numerical framework for near-field optics}, Proceedings of the Royal Society A, \textbf{474} (2018), 20180150. 


\bibitem{Ammari-zhang18-1}
H. Ammari, M. Ruiz, S. Yu, and H. Zhang,
{\em Reconstructing fine details of small objects by using
plasmonic spectroscopic data}, SIAM J. Imaging Sci., \textbf{11} (2018), 1-23.

\bibitem{Ammari-zhang18-2}
H. Ammari, M. Ruiz, S. Yu, and H. Zhang,
{\em Reconstructing fine details of small objects by using
plasmonic spectroscopic data. Part II: The Strong Interaction Regime}, SIAM J. Imaging Sci., \textbf{11} (2018), 1931-1953.

\bibitem{Anker2008}
J. N. Anker, W. P. Hall, O. Lyandres, N. C. Shah, J. Zhao, and R. P. Van Duyne, 
{\em Biosensing
with plasmonic nanosensors}, Nature Material, \textbf{7} (2008), 442-453.

\bibitem{Blanchard-Meunier}
A. Blanchard-Dionne and M. Meunier, {\em Sensing with periodic nanohole arrays}, Advances in Optics and Photonics, \textbf{9} (2017). 891-940.

\bibitem{Brolo}
A. Brolo, R. Gordon, B. Leathem, and K. L. Kavanagh, {\em Surface plasmon sensor based on the enhanced light transmission through arrays of nanoholes in gold films}, 
Langmuir, \textbf{20} (2004). 4813--4815.

\bibitem{Cetin}
A. Cetin,  \textit{et al.}, {\em Plasmonic nanohole arrays on a robust hybrid substrate for highly sensitive label-free biosensing}, ACS Photonics, \textbf{2} (2015), 1167-1174.

\bibitem{Costabel-2007}
M. Costabel, M. Dauge and R. Duduchava { \em Asymptotics Without Logarithmic Terms for Crack Problems},  Communications in Partial Differential Equations,
\textbf{28} (2003), 869-926.

\bibitem{Dahlin}
A. Dahlin,  N. Wittenberg, F. H\"{o}\"{o}k, and S. H., Oh, {\em Promises and challenges of nanoplasmonic devices for refractometric biosensing}, Nanophotonics, \textbf{2} (2013), 83-101.

\bibitem{Dhawan}
A. Dhawan, M. Gerhold, and J. Muth, {\em Plasmonic structures based on subwavelength apertures for chemical and biological sensing applications},
IEEE Sensors Journal \textbf{8} (2008): 942-950.

\bibitem{Ebbesen-1998}
T. W. Ebbesen, H. J. Lezec, H. F. Ghaemi, T. Thio, and P. A. Wolff, { \em Extraordinary
optical transmission through sub-wavelength hole arrays}, Nature, \textbf{391} (1998), 667-669.


\bibitem{Vidal-2010}
F. J. Garcia-Vidal, L. Martin-Moreno, T. W. Ebbesen, and L. Kuipers, {\em Light passing
through subwavelength apertures}, Rev. Modern Phys., \textbf{82} (2010), 729-787.

\bibitem{Gome-Crutz}
J. Gomez-Cruz, \textit{et al.} {\em Cost-effective flow-through nanohole array-based biosensing platform for the label-free detection of uropathogenic E. coli in real time},
Biosensors and Bioelectronics, \textbf{106} (2018), 105-110.

\bibitem{Lee}
S. H. Lee, \textit{et al.}, {\em Linewidth optimized extraordinary optical transmission in water with template stripped metallic nanohole arrays},
Advanced Functional Materials, \textbf{22} (2012), 4439-4446.

\bibitem{Li}
X. Li, \textit{et al.},  {\em Plasmonic nanohole array biosensor for label-free and real-time analysis of live cell secretion}, Lab on a Chip, \textbf{17} (2017), 2208-2217.

\bibitem{LONR}
J. Lin, S-H. Oh, H-M. Nguyen, and F. Reitich, {\em Field enhancement and saturation of millimeter waves inside a metallic nanogap}, Opt. Express, \textbf{22} (2014), 14402-14410.

\bibitem{LR}
J. Lin and F. Reitich, {\em Electromagnetic field enhancement in small gaps: a rigorous mathematical theory}, SIAM J. Appl. Math., \textbf{75} (2015), 2290-2310.

\bibitem{lin_shipman_zhang}
J. Lin, S. Shipman, and H. Zhang,  {\em A mathematical theory for Fano resonance in a periodic array of narrow slits}, SIAM J. Appl. Math., to appear.

\bibitem{lin_zhang17}
J. Lin and H. Zhang, {\em Scattering and field enhancement of a perfect conducting narrow slit}, SIAM J. Appl. Math.,  \textbf{77} (2017), 951--976.

\bibitem{lin_zhang18_1}
J. Lin and H. Zhang, {\em Scattering by a periodic array of subwavelength slits I: field enhancement in the diffraction regime}, Multiscale Model. Simul., \textbf{16} (2018), 922--953.

\bibitem{lin_zhang18_2}
J. Lin and H. Zhang, {\em Scattering by a periodic array of subwavelength slits II: surface bound states, total transmission and field enhancement in the homogenization regimes},
 Multiscale Model. Simul., \textbf{16} (2018), 954--990.

\bibitem{lin_zhang19_1}
J. Lin and H. Zhang, {\em  An integral equation method for numerical computation of scattering resonances in a narrow metallic slit},
J. Comput. Phys., \textbf{385} (2019), 75-105.

\bibitem{lin_zhang19_2}
J. Lin and H. Zhang, {\em Mathematical analysis of surface plasmon resonance by a nano-gap in the plasmonic metal},
SIAM Math. Anal., \textbf{51} (2019), 4448-4489.

\bibitem{lin_zhang19_3}
J. Lin and H. Zhang, {\em Mathematical theory of anomalous scattering by a subwavelength slit in a plasmonic metallic slab},
to be submitted. 

\bibitem{Oh_Altug}
S.-H. Oh and H. Altug, {\em Performance metrics and enabling technologies for nanoplasmonic biosensors}, Nat. Commun., \textbf{9} (2018), 5263.

\bibitem{PHSF}
L. Pang, G. M. Hwang, B. Slutsky, and Y. Fainman, {\em Spectral sensitivity of two-dimensional nanohole array surface plasmon polariton resonance sensor},
Appl. Phys. Lett., \textbf{91} (2007), 123112.

\bibitem{Rodrigo}
S. Rodrigo, F. de León-Pérez, and L. Mart\'{i}n-Moreno, {\em Extraordinary optical transmission: fundamentals and applications}, Proceedings of the IEEE, \textbf{104} (2016), 2288-2306.

\bibitem{Willets2007}
K. A. Willets and R. P. Van Duyne, 
{\em Localized surface plasmon resonance spectroscopy and sensing},
Annu. Rev. Phys. Chem. \textbf{58} (2007), 267-297.


\bibitem{Soler}
M. Soler, \textit{et al.} {\em Multiplexed nanoplasmonic biosensor for one-step simultaneous detection of Chlamydia trachomatis and Neisseria gonorrhoeae in urine},
Biosensors and Bioelectronics, \textbf{94} (2017), 560-567.

\end{thebibliography}
\end{document}